\documentclass{emulateapj}
\usepackage{txfonts}
\pdfoutput=1

\newcommand{\kB}{\ensuremath{k_\mathrm{B}}} 
\newcommand{\NA}{\ensuremath{N_\mathrm{\!A}}} 
\newcommand{\mb}{\ensuremath{m_\mathrm{u}}} 

\newcommand{\unitspace}{\ensuremath{\,}}
\newcommand{\usp}{\unitspace}
\newcommand{\numberspace}{\ensuremath{\;}}
\newcommand{\nsp}{\numberspace}
\newcommand{\unitstyle}[1]{\ensuremath{\mathrm{#1}}}
\newcommand{\power}[2]{\ensuremath{{#1}^{#2}}}

\newcommand{\kilo}{\unitstyle{k}}
\newcommand{\Mega}{\unitstyle{M}}

\newcommand{\cm}{\unitstyle{cm}}
\newcommand{\gram}{\unitstyle{g}}

\newcommand{\meter}{\unitstyle{m}}

\newcommand{\second}{\unitstyle{s}}

\newcommand{\Kelvin}{\unitstyle{K}}
\newcommand{\K}{\Kelvin}  

\newcommand{\grampercc}{\gram\usp\power{\cm}{-3}} 
\newcommand{\grampersquarecm}{\gram\usp\power{\cm}{-2}} 

\newcommand{\columnunit}{\grampersquarecm}
\newcommand{\erg}{\unitstyle{ergs}} 
\newcommand{\ergs}{\erg}
\newcommand{\ergspersecond}{\erg\unitspace\power{\second}{-1}}

\newcommand{\fermi}{\unitstyle{fm}} 
\newcommand{\eV}{\unitstyle{eV}}        
\newcommand{\keV}{\kilo\eV} 
\newcommand{\MeV}{\Mega\eV} 

\newcommand{\Msun}{\ensuremath{M_\odot}}

\newcommand{\yr}{\unitstyle{yr}}        
\newcommand{\km}{\kilo\meter}   

\newcommand{\dif}{\ensuremath{\mathrm{d}}}
\newcommand{\ee}[1]{\ensuremath{\times 10^{#1}}}

\newcommand{\satellite}[1]{\emph{#1}}

\newcommand{\rxte}{\satellite{RXTE}}

\newcommand{\nuclei}[2]{\ensuremath{\mathrm{^{#1}#2}}}
\newcommand{\helium}[1][4]{\nuclei{#1}{He}}
\newcommand{\carbon}[1][12]{\nuclei{#1}{C}}
\newcommand{\oxygen}[1][16]{\nuclei{#1}{O}}
\newcommand{\iron}[1][56]{\nuclei{#1}{Fe}}

\newcommand{\source}[3]{#1~#2$#3$}
\newcommand{\ks}{\source{KS}{1731}{-260}}
\newcommand{\mx}{\source{MXB}{1659}{-29}}
\newcommand{\axj}{\source{AX}{J1754.2}{-2754}}
\newcommand{\Ye}{\ensuremath{Y_{e}}}
\newcommand{\YN}{\ensuremath{Y_{N}}}

\newcommand{\grampersecond}{\gram\usp\power{\second}{-1}}

\newcommand{\Qimp}{\ensuremath{Q_{\mathrm{imp}}}}
\newcommand{\qnuc}{\ensuremath{\epsilon_{\mathrm{nuc}}}}
\newcommand{\qnu}{\ensuremath{\epsilon_{\nu}}}
\newcommand{\ephi}{\ensuremath{e^{\phi/c^{2}}}}
\newcommand{\etphi}{\ensuremath{e^{2\phi/c^{2}}}}
\newcommand{\ytop}{\ensuremath{y_{\mathrm{top}}}}
\newcommand{\Teff}{\ensuremath{T_{\mathrm{eff}}}}
\newcommand{\Teffinf}{\ensuremath{\Teff^{\infty}}}
\newcommand{\Mdot}{\ensuremath{\dot{M}}}
\newcommand{\unitday}{\unitstyle{d}}

\begin{document}
\title{Mapping crustal heating with the cooling lightcurves of quasi-persistent transients}
\author{Edward F. Brown}
\affil{Department of Physics \& Astronomy, National
Superconducting Cyclotron Laboratory, and the Joint Institute for
Nuclear Astrophysics, Michigan State University, East  Lansing, MI 48824}
\email{ebrown@pa.msu.edu}
\and
\author{Andrew Cumming}
\affil{Department of Physics, McGill University, 3600 rue University, Montreal, QC, H3A2T8, Canada}
\email{cumming@physics.mcgill.ca}
\submitted{to appear in The Astrophysical Journal}

\begin{abstract}
The monitoring of quiescent emission from neutron star transients with accretion outbursts long enough to significantly heat the neutron star crust has opened a new vista onto the physics of dense matter. In this paper we construct models of the thermal relaxation of the neutron star crust following the end of a protracted accretion outburst. We confirm the finding of Shternin et al., that the thermal conductivity of the neutron star crust is high, consistent with a low impurity parameter. We describe the basic physics that sets the broken power-law form of the cooling lightcurve. The initial power law decay gives a direct measure of the temperature profile, and hence the thermal flux during outburst, in the outer crust. The time of the break, at hundreds of days post-outburst, corresponds to the thermal time where the solid transitions from a classical to quantum crystal, close to neutron drip. We calculate in detail the constraints on the crust parameters of both \ks\ and \mx\ from fitting their cooling lightcurves. Our fits to the lightcurves require that the neutrons do not contribute significantly to the heat capacity in the inner crust, and provide evidence in favor of the existence of a neutron superfluid throughout the inner crust. Our fits to both sources indicate an impurity parameter of order unity in the inner crust.
\end{abstract}

\keywords{ dense matter --- stars: neutron --- X-rays: binaries --- X-rays: individual(\ks, \mx) }

\section{Introduction}\label{s.introduction}

In 2001 the low mass X-ray binary \ks\ went into quiescence \citep{wijnands:ks1731} after accreting steadily since its first detection about 12 years prior \citep{sunyaev:transient}. \cite{rutledge.ea.01:ks1731} suggested that the neutron star crust would be heated out of thermal equilibrium with the core during this long outburst, and that monitoring observations could detect the thermal relaxation of the crust following the cessation of accretion. A regular monitoring program of \ks\ with Chandra and XMM \citep{wijnands:ks1731,wijnands.ea:xmm_1731,Cackett2006Cooling-of-the-} has indeed detected a steadily decaying luminosity. Another source, \mx, has also been observed to cool following a 2.5~yr-long outburst \citep{wijnands03:mxb1659-298,wijnands.homan.ea:monitoring}, and recent observations show that the cooling appears to have halted \citep{Cackett2008Cooling-of-the-}. 
Recently a third source, \axj, was observed to turn off \citep{Bassa2008The-faint-neutr} after being in outburst since being detected in 1999 \citep{Sakano2002ASCA-X-Ray-Sour}. This source is likely an ultracompact binary, and is of particular interest as it may be similar to \source{1H}{1905}{+000}, which has an extremely low quiescent flux, perhaps indicating a very low neutron star core temperature \citep{Jonker2006The-neutron-sta}. Finally, the rapidly accreting source \source{XTE}{J1701}{-462} went into quiescence in 2007 after $\approx 1.6\nsp\yr$ of active accretion \citep{Homan2007Rossi-X-Ray-Tim,Altamirano2007Following-the-n}, and \source{EXO}{0748}{-676} entered quiescence in 2008 after being in outburst for 24\nsp\yr\ \citep{Degenaar2008Chandra-and-Swi}.

\cite{rutledge.ea.01:ks1731} emphasized that observations of the crust cooling would offer a new probe of the crust thermal properties. The crust of a neutron star in a LMXB may have significantly different properties than that of a young isolated neutron star, as the accretion lifetime of a LMXB is long enough for accreted matter to replace the entire crust. The accreted crust is heated during outburst by electron captures and pycnonuclear reactions, mostly at densities close to neutron drip \citep{haensel90a,haensel.zdunik:nuclear,Haensel2008Models-of-crust}. The temperature profile in the crust is determined by how the heat released in this ``deep crustal heating'' is transported outwards to the surface or inwards to the neutron star core, where it is radiated as neutrinos. 

A particularly uncertain property of accreted crusts is the thermal conductivity. The matter entering the top of the crust is expected to consist of a mixture of elements produced by hydrogen and helium burning at low densities. \citet{schatz99} found that rp-process burning produced ashes with $\Qimp\approx 100$, where the impurity parameter 
\begin{equation}
\Qimp \equiv n_{\mathrm{ion}}^{-1}\sum_{i}n_{i}(Z_{i}-\langle Z\rangle)^{2}
\end{equation}
measures the distribution of nuclide charge numbers, an important parameter for setting the conductivity of the crust \citep{itoh93}. \citet{brown:nuclear} suggested that at such large values of $\Qimp$, the crust would have an amorphous structure, with a low thermal conductivity comparable to the conductivity of the liquid state. Recent molecular dynamics calculations of the solidification of such a mixture \citep{Horowitz2007Phase-Separatio,Horowitz2008Thermal-conduct} show, however, that a regular crystal structure does form, but that phase separation between the liquid and solid phases or the formation of multiple solid phases can occur, changing the effective value of \Qimp\ in the crust compared to the rp-process mixture. In addition, nuclear reactions in the crust may also act to reduce the effective value of \Qimp\ \citep{Horowitz2008Thermal-conduct}. Measurement of $\Qimp$ in the crust would therefore give an important test of the composition of rp-process ashes and their subsequent evolution as they are compressed towards nuclear density.

\citet{Shternin2007Neutron-star-co} compared time-dependent calculations of deep crustal heating and subsequent cooling to the observations of \ks. They found that a low value of thermal conductivity corresponding to an amorphous crust could be ruled out because it would give cooling on a much longer timescale than observed. In this paper, we confirm this result, and go further by describing the basic physics that sets the shape of the cooling lightcurve, and calculating in detail the constraints on $\Qimp$ and other crust parameters coming from the cooling lightcurves of both \ks\ and \mx. \citet{Cackett2006Cooling-of-the-} found that both of these decays could be fit with an exponential decay to a constant, although a single power-law ($L\propto t^{-\alpha}$, with $\alpha = 0.50\pm0.03$) also adequately fits the data for \ks\ \citep{Cackett2008Cooling-of-the-}. We show here that the lightcurve of a cooling crust is a broken power law. The initial power law decay gives a direct measure of the temperature profile, and hence the thermal flux during outburst, in the outer crust. The time of the break, at hundreds of days post-outburst, corresponds to the thermal time where the solid transitions from a classical to quantum crystal, close to neutron drip. At late times, the luminosity levels off at a value set by the neutron star core temperature.

We start in \S 2 by describing our time-dependent cooling calculations and an analytic model of the results, and go on in \S 3 to calculate the constraints on crust parameters coming from comparison with the observed cooling of \ks\ and \mx. The Appendix discusses the details of our crust models.

\section{Models of Crust Cooling in \ks\ and \mx}\label{s.crust-models}

\subsection{Hydrostatic structure of the crust}\label{s.hydrostatic-structure}

Because the temperature is always low relative to the electron and neutron Fermi energies, we can solve for the temperature and luminosity using a hydrostatic structure. In the crust, the pressure $P$ makes a convenient Eulerian coordinate, and we integrate the equations \citep{thorne77}  for the radius $r$, gravitational mass $M$, and potential $\phi$,
\begin{eqnarray}
\label{e.dR}\frac{\dif r}{\dif \ln P}  &=& -\frac{P}{\rho g}(1+z)^{-1},\\
\label{e.dM}\frac{\dif M}{\dif \ln P} &=& -4\pi r^{2}\frac{P}{g},\\
\label{e.dPhi}\frac{\dif \phi}{\dif \ln P}  &=& -\frac{P}{\rho}.
\end{eqnarray}
Here $1+z = [1-2GM/(rc^{2})]^{-1/2}$, $g = GM(1+z)/r^{2}$ is the gravitational acceleration, and $\rho$ is the density of mass-energy. We have neglected terms $\mathcal{O}(pr^{3}/Mc^{2})$, as appropriate in the crust. As boundary conditions, we assume a transition density to uniform $npe$ matter at $n = 0.08\nsp\fermi^{-3}$ (consistent with recent studies of clustering in uniform nuclear matter; \citealt{Oyamatsu2007Symmetry-energy}), and set $M$ and $r$ according to a neutron star model computed using the EOS of \citet{akmal98}.  We integrate outwards to a pressure $P=2.3\ee{26}\nsp\ergs\usp\cm^{-3}$, corresponding to a column depth from the surface\footnote{In a thin layer, the column depth is $\int_{r}^{\infty}\;\rho\,\dif r \approx P/g$; in this paper we will use the term to refer to $y\equiv P/g$.} of $P/g = 10^{12}\nsp\gram\usp\cm^{-2}$, at which point we apply the third boundary condition $\phi(r=R) = (c^{2}/2)\ln[1-2GM/(Rc^{2})]$. The integration is performed using a standard fourth-order Runge-Kutta algorithm, and the output is constrained to generate points uniformly distributed in $\ln P$ for use in the time-dependent code (\S~\ref{s.time-dependent-model}). Our equation of state, as well as our model for the composition, is detailed in the Appendix.

\begin{figure}[htbp]
\includegraphics[width=\columnwidth]{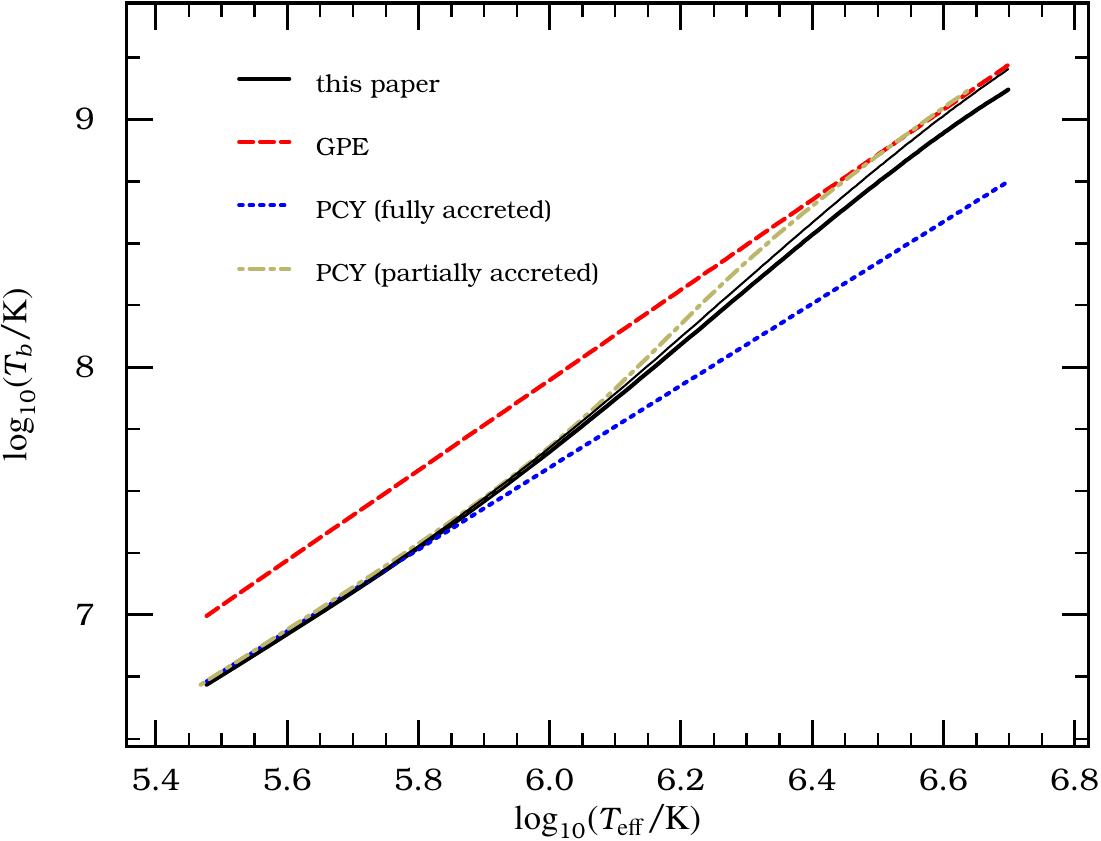}
\caption{\label{f.teff-tb}Temperature in the neutron star ocean ($T_{b}$) as a function of photosphere temperature $\Teff$ (\emph{solid line}).  The relation of \protect\citet[\emph{dashed line}]{gudmundsson83} is shown for comparison. We also show two models from \citet{potekhin97}: their ``fully accreted'' model \emph{dotted line} and a ``partially accreted'' model (\emph{dot-dashed line}) in which the light elements are in the region $P/g < 10^{9}\nsp\columnunit$. Note that for these latter two models, the temperature $T_{b}$ is taken at a density $10^{10}\nsp\grampercc$ ($P/g \approx 4.3\ee{13}\nsp\columnunit$), which is somewhat deeper than the boundary used in our calculations. For comparison with these cases, we also show (\emph{thin solid line}) our relation obtained by integrating to this depth.}
\end{figure}

\subsection{Time-dependent Heating and Cooling}\label{s.time-dependent-model}

The time-dependent equations for the evolution of temperature and luminosity are
\begin{eqnarray}\label{e.temperature}
\frac{\partial}{\partial t}\left(T \ephi\right) &=& \etphi\frac{\qnuc-\qnu}{C} - \frac{1}{4\pi r^{2}\rho C (1+z)} \frac{\partial}{\partial r}\left(L\etphi\right)\\
\label{e.luminosity}
L\etphi &=& -\frac{4\pi r^{2} K\ephi}{1+z}\frac{\partial}{\partial r}\left(T\ephi\right),
\end{eqnarray}
where \qnuc\ and \qnu\ are the specific nuclear heating and neutrino emissivity, $C$ is the specific heat, and $K$ is the thermal conductivity. We solve these equations using the method of lines. We use the common technique of defining $L\etphi$ at the midpoints of our grid by interpolating $4\pi r^{2}K\ephi/(1+z)$ and differencing $T\ephi$; as a result the divergence term in equation~(\ref{e.temperature}) is second-order and explicitly conserves flux. This procedure yields a set of coupled ordinary differential equations, which we then integrate using a semi-implicit extrapolation method \citep[see][and references therein]{pre92}. Our calculation of $C$, $K$, 
\qnuc, and \qnu\ is described in the Appendix.

We used two different boundary conditions for the core.  The first is to simply assume a constant temperature, which we fit to observations. The second is to match the inwards luminosity at the crust-core boundary to the neutrino emission from the core using a tabulated $T_c$-$L_{\nu}$ relation for different assumptions of the core neutrino emissivity. In this way, we self-consistently solve for the core temperature appropriate for the assumed core physics rather than treat it as a free parameter. Unless the quiescent interval is long, we find that the core temperature is essentially constant over an outburst-quiescence cycle.

The boundary condition at the surface is more ambiguous.  During an outburst, the temperature in the neutron star envelope is set by the burning of hydrogen and helium, and (possibly) fusion of light elements such as \carbon.  Our code does not track this burning, and so we fix the temperature at $P/g = 10^{12}\nsp\columnunit$ at a fixed value $T_{b}$. This column is roughly where superburst ignition occurs, and should demarcate the bottom of the region containing light element, unstable reactions.  Because our first goal is to fit the lightcurve, for now we do not set $T_{b}$ to any expected value \emph{a priori}, but rather leave it as an adjustable parameter.

During quiescence, we calculate the cooling flux at the top of the grid using a tabulated relation between \Teffinf\ and the temperature obtained by integrating the steady-state thermal structure of the neutron star envelope \citep{brown.bildsten.ea:variability}. In these integrations, we fix the envelope to be pure \helium\ down to a depth $P/g = 10^{9}\nsp\columnunit$, with a layer of pure \iron\ down to a depth $P/g = 10^{12}\nsp\columnunit$. The resulting relation (Fig.~\ref{f.teff-tb}, \emph{solid line}) resembles that of \citet[\emph{dashed line}]{gudmundsson83} at low $\Teff$, but trends towards the fully accreted model of \citet[\emph{dotted line}]{potekhin97} at higher \Teff. We also show the calculation of \citep{potekhin97} for a light element envelope that is a accreted to a density (\emph{dot-dashed line}). Note that our boundary is not quite as deep as the one used by \citet{gudmundsson83} or \citet{potekhin97}.  For comparison, we therefore show (\emph{thin black line}) our relation $T_{b}(\Teff)$ when the bottom of the \iron\ layer is taken at the same density as used by \citet{gudmundsson83} and \citet{potekhin97}.

\subsection{Numerical Results}\label{s.numerical}

We now describe our method for fitting the data, using \mx\ as an example. With the hydrostatic structure constructed (eqs.~[\ref{e.dR}]--[\ref{e.dPhi}]), we use the surface gravity and redshift to tabulate $T_{b}(\Teffinf)$, where $T_{b}$ is the temperature at $y=\ytop$. We set the core temperature from the last observation of \mx\ \citep{Cackett2008Cooling-of-the-}. Although our code can follow the thermal evolution of the core, the change in temperature is not large unless the quiescent interval is $\sim 10^{3}\nsp\yr$, the core cooling timescale.  We therefore found it more convenient to fix the core temperature at this value. Table~\ref{t.parameters} shows the parameters for the calculations---accretion rate, core temperature, outburst duration, and recurrence time---described in this section.  In both cases, we used a neutron star with a mass $1.62\nsp\Msun$, radius $11.2\nsp\km$, surface gravity $2.27\ee{14}\nsp\cm\usp\second^{-2}$, and surface redshift 1.32.

\begin{deluxetable}{cccc}
\tablecaption{Parameters for numerical integrations}
\tablewidth{\columnwidth}
\tablehead{\colhead{quantity} & \colhead{unit} & \colhead{\ks} & \colhead{\mx}}
\startdata
$\dot{M}$ & \grampersecond & $10^{17}$ & $10^{17}$\\
$T_{\mathrm{core}}^{\infty}$ & $10^{7}\nsp\K$ & 4.6  & 2.6 \\
outburst duration & \yr & 12.0 & 2.5\\
recurrence time & \yr & 100 & 21\\
\enddata
\label{t.parameters}
\end{deluxetable}

We base the outburst and recurrence times on observations. \ks\ was first discovered in August 1989 with \satellite{Mir}/Kvant \citep{sunyaev:transient} and went into quiescence in 2001 \citep{wijnands:ks1731}. Its recurrence time is unknown, although it has been quiescent since then.  So long as the recurrence time is longer than the thermal relaxation time of the crust (as determined by \citealt{Shternin2007Neutron-star-co} and confirmed here), our results are not sensitive to the recurrence time since we are using the core temperature as a parameter in our fit. \mx\ has had two outbursts since being discovered in 1976 \citep{Lewin1976X-Ray-Bursts}. It was initially detected with \satellite{SAS3} and \satellite{HEAO} through 1978; subsequent observations with a variety of instruments \citep[see][and references therein]{wijnands03:mxb1659-298} failed to detect the source until it was detected in outburst again in 1999 \citep{in-t-Zand1999V2134-Ophiuchi-}. This outburst was monitored with \rxte/ASM until the source went into quiescence in September 2001. We adopt an outburst duration of 2.5\nsp\yr\ and a recurrence time of 21\nsp\yr. For both sources we adopt an accretion rate $\Mdot = 10^{17}\nsp\grampersecond$, which is consistent with recent estimates \citep[][see \S~\ref{s.constraints-mdot}]{Galloway2008Thermonuclear-t}, for example from the X-ray luminosity using the relation for energy radiated by radially infalling material \citep[see, e.~g.,][]{ayasli82}  $\Mdot = (1+z)L^{\infty}/c^2 z$. There is uncertainty in deriving the mass accretion rate from the source luminosity, and we shall explore the sensitivity of our conclusions to the assumed mass accretion rate in \S~\ref{s.constraints-mdot}.

\begin{figure}[htbp]
\includegraphics[width=\columnwidth]{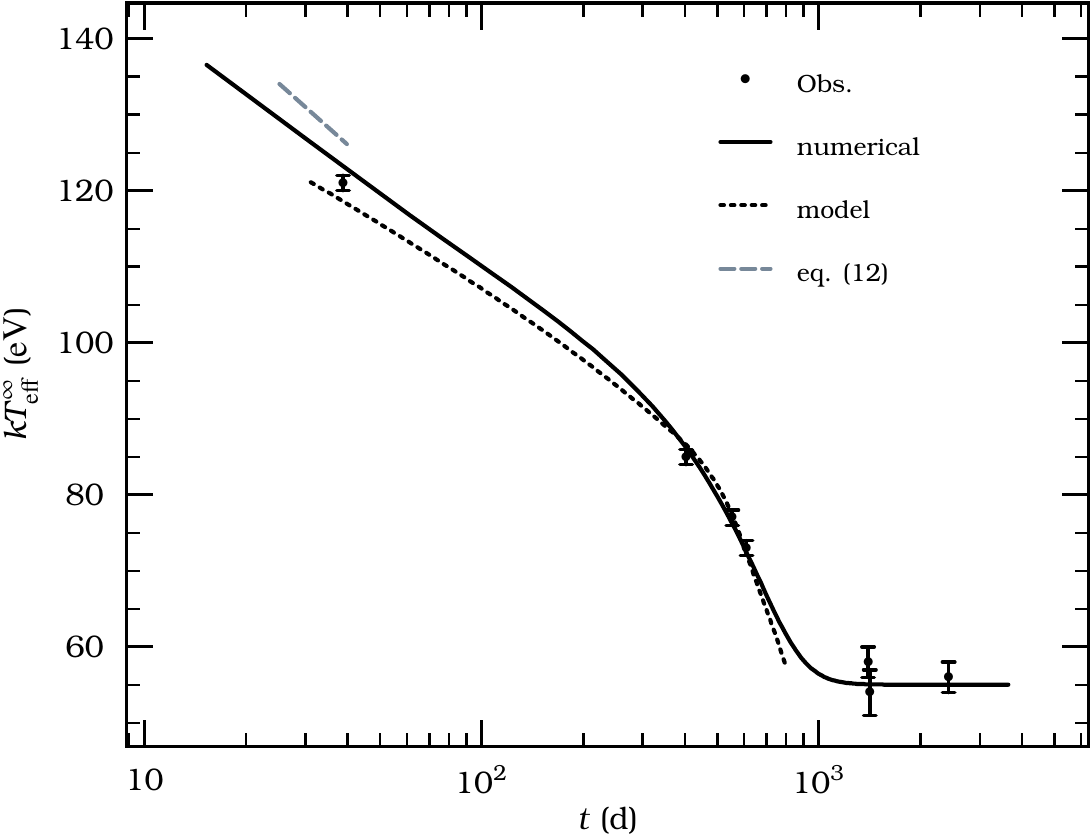}
\caption{An example of a lightcurve that fits the observed cooling of \mx. The numerical model is shown as a solid curve and has $\Qimp=4.0$ and $T_b=3.8\ee{8}\nsp\K$. The dotted curve shows the corresponding toy model lightcurve from \S \ref{s.simple-model}, and we also show (\emph{grey dashed line}) the slope given by eq.~(\ref{e.slope-Teff}).
\label{f.mxb1659-model}}
\end{figure}

\begin{figure}[htbp]
\includegraphics[width=\columnwidth]{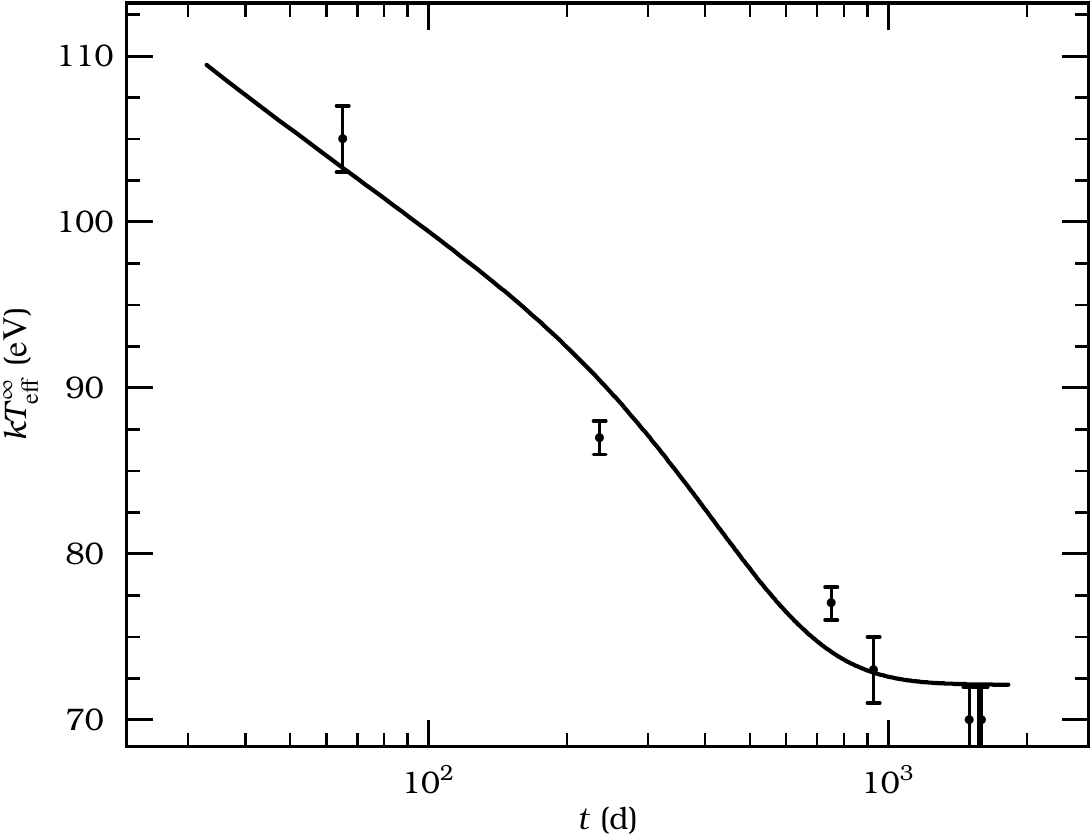}
\caption{A model lightcurve for \ks, with $\Qimp=1.5$ and $T_{b}=2.5\ee{8}\nsp\K$.\label{f.ks-lightcurve}}
\end{figure}

To generate a starting model, we first compute the thermal structure for a steadily accreting neutron star at the outburst accretion rate $\Mdot $.  We then turn on the time-dependent terms in equations~(\ref{e.temperature})--(\ref{e.luminosity}). We run through several outburst/quiescent cycles, and then generate a quiescent lightcurve with finer time resolution. Because the recurrence time is unknown for \ks, we use in the code an arbitrary recurrence time of 100\nsp\yr; note that our results do not depend on the value of the recurrence time, however, since the core temperature is held fixed.
The code stores snapshots of the temperature during this ``high resolution'' run so that the physical properties of the crust during its gloaming can be reconstructed.

Several runs with different values of \Qimp\ and $T_{b}$ were then made, adjusting the chosen values to provide a good fit by eye to the lightcurve.  Although there is no physical reason why \Qimp\ should be the same value throughout the crust, there are no reliable calculations of how \Qimp\ would evolve with depth, and we therefore choose to set \Qimp\ to a constant. Indeed, we shall show (\S~\ref{s.simple-model}) that it is really the value of \Qimp\ in the inner crust that is important for determining the lightcurve. Figure \ref{f.mxb1659-model} shows a fit for \mx, with $\Qimp=4.0$ and $T_{b} = 3.8\ee{8}\nsp\K$. The lightcurve is a broken power law with a break at $t\approx 300\nsp\unitstyle{d}$, going to a constant at late times. 

Repeating this procedure for \ks, we find an acceptable fit with $\Qimp=1.5$, $T_{b} = 2.5\ee{8}\nsp\K$, as shown in Fig.~\ref{f.ks-lightcurve}. Note that the errors on the temperature given by \citet{Cackett2006Cooling-of-the-} are (erroneously) stated to be $1\textrm{-}\sigma$ errors, but are in fact 90\% confidence limits \citep{Cackett2008Cooling-of-the-}. We show $1\textrm{-}\sigma$ errors in all figures in this paper.  For the observations of \mx\ we use the published $1\textrm{-}\sigma$ errors \citep{Cackett2008Cooling-of-the-}; for the  observations of \ks\, we assume a Gaussian distribution to adjust the published 90\% confidence limits \citep{Cackett2006Cooling-of-the-} to $1\textrm{-}\sigma$ errors.

\subsection{Simple understanding of the lightcurve}\label{s.simple-model}

We now describe a simple model that accurately reproduces the lightcurve from the time-dependent calculation, and reveals the basic physics underlying the lightcurve. This will allow us to understand the effect of different parameters on the lightcurve. A similar approach has been applied to lightcurves of white dwarfs cooling following a dwarf nova \citep*{Piro2005White-Dwarf-Hea}, and the early phase of superburst lightcurves (Cumming et al.~2006, Appendix A).

\begin{figure}[htbp]
\includegraphics[width=\columnwidth]{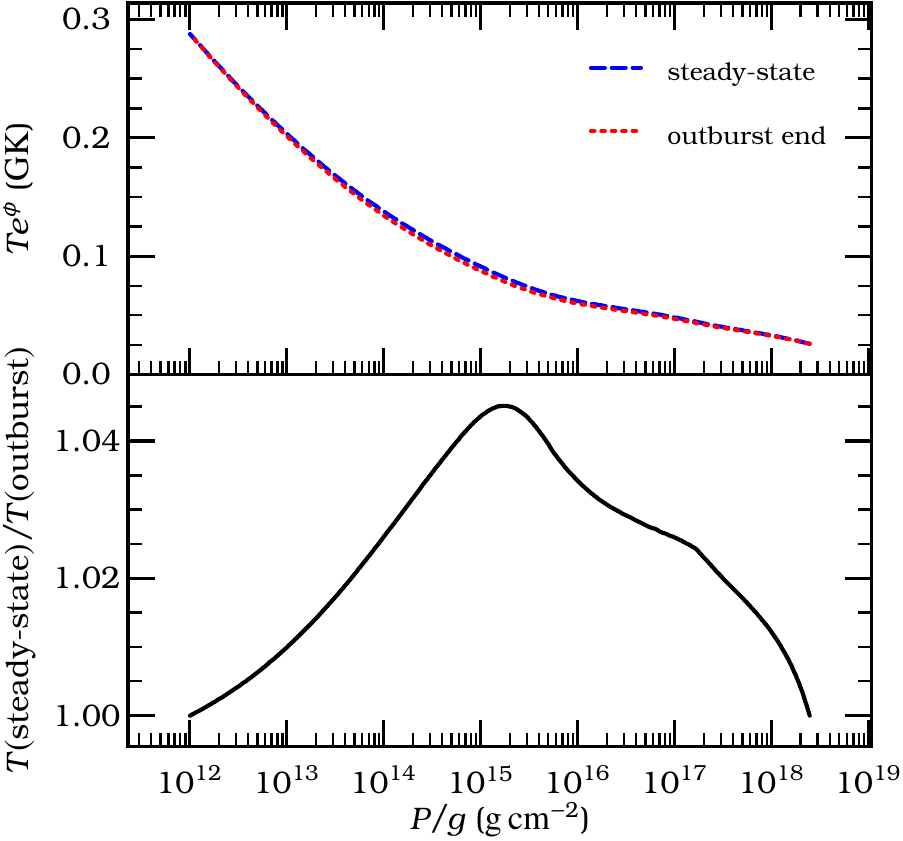}
\caption{(\emph{Top panel}) Temperature in the crust of \mx\ at the end of a 2.5\nsp\yr\ outburst (\emph{dotted line}) and that of a neutron star in thermal steady-state accreting at the outburst accretion rate (\emph{dashed line}). Note that the core temperature and the temperature at a column $\ytop=10^{12}\nsp\columnunit$ are fixed.  (\emph{Bottom panel}) Ratio of the steady-state temperature to that at the outburst end.\label{f.mxb-Tcompare}}
\end{figure}

We first note that during the long outbursts of \mx\ and \ks, the crust temperature profile is very close to the thermal steady-state profile at the outburst accretion rate. This is shown in Fig.~\ref{f.mxb-Tcompare} for \mx, in which we compare the temperature at the end of the outburst of \mx\ in our numerical calculations (\emph{top panel, dotted line}) with the temperature profile of a steady-state calculation at the outburst accretion rate (\emph{top panel, dashed line}). The largest difference (Fig.~\ref{f.mxb-Tcompare}, \emph{bottom panel}), in the inner crust where the strongest heat sources lie, is only 4\% percent. Therefore a good approximation to the initial temperature profile for the cooling is the steady-state profile. This is an even better approximation for \ks, which had a longer outburst than \mx. Our calculation is therefore different from that of \citet{ushomirsky.rutledge:time-variable}, who injected the entire heat deposition of the outburst at once.

\begin{figure}[htbp]
\includegraphics[width=\columnwidth]{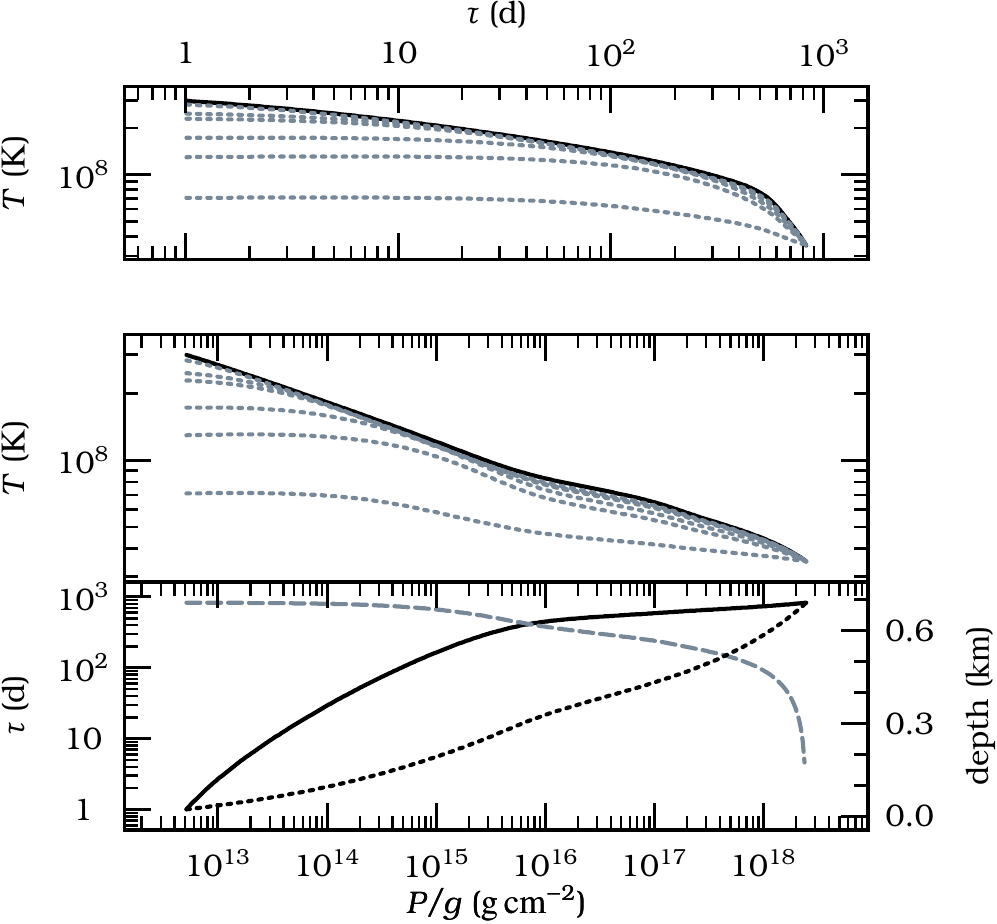}
\caption{\label{f.T-tau} Illustration of the cooling behavior of our analytical model. The bottom panel shows the thermal diffusion time, as a function of $P/g$, to the surface (eq.~[\ref{e.analytical-tau-outer}]; \emph{solid line})  and to the core (\emph{grey dashed line}), as well as the depth from the surface (\emph{dotted black line, right axis}). The middle panel shows the temperature as a function of $P/g$.  In the top panel, we plot the temperature (\emph{solid line}) against $\tau$ at the start of the quiescent phase. Subsequent ``snapshots'' of the temperature in the crust (\emph{grey dotted lines}) at $t = 3.1$, 10, 31, 107, 305, and 504\nsp\unitday\ plotted against $\tau$ (\emph{top panel}) and $P/g$ (\emph{middle panel}).}
\end{figure}

Starting with the initial temperature profile, we can understand the evolution of the cooling layer and the resulting lightcurve by noting that at a given depth, the thermal evolution occurs on the characteristic thermal timescale associated with that depth. This is illustrated in the middle panel of Figure \ref{f.T-tau}, which shows snapshots of the temperature profile of the crust as it cools during quiescence. At a given time, the temperature profile has two parts: the inner layers have not yet started to cool and still have the temperature profile corresponding to the initial condition (the steady state profile during outburst); the outer layers have relaxed thermally and the temperature profile there corresponds to a constant outwards flux. The transition occurs at a depth where the thermal time at that depth is equal to the current time. In the bottom panel of Figure \ref{f.T-tau}, we show the thermal time as a function of depth, where we calculate the thermal time from the surface following \cite{henyey69}, \begin{equation}\label{e.tau}
\tau \equiv \frac{1}{4}\left[\int_{0}^{z}\left(\frac{\rho C_{P}}{K}\right)^{1/2} dz'\right]^{2}.
\end{equation}
where $\rho$ is the density, $C_{P}$ the specific heat, and $K$ the thermal conductivity. In the top panel of Figure \ref{f.T-tau} we show the temperature profiles as a function of the thermal time. This shows directly that the deviation of each of the dashed temperature profiles away from the initial temperature profile occurs at a depth where the thermal time is approximately equal to the time since cooling began.

The temperature of the inner crust is also affected by conduction of heat into the core. We show the timescale for thermal diffusion into the core in Figure \ref{f.T-tau} (\emph{bottom panel, dotted line}). The two thermal times intersect at a depth ($P/g\sim 10^{16}\nsp\columnunit$) where the thermal diffusion time is $\approx 400\nsp\unitday$. After this point the temperature in the inner crust drops markedly (\emph{top panel}).

This understanding suggests a simple model of the lightcurve. We start with the initial temperature profile set by the steady-state profile at the outburst accretion rate. Then, for each time $t$, we locate the depth at which $\tau=t$. We then find the outwards flux in a constant flux solution that has a temperature equal to the initial temperature at the depth where $t=\tau$. This value of flux is the flux emerging from the surface at time $t$. The dotted curve in Figure \ref{f.mxb1659-model} shows a lightcurve calculated in this way, using the same parameters $\Qimp$, $T_b$, and $T_c$ as the numerical model. The simple model shows excellent agreement with the numerical model. 

\begin{figure}[htbp]
\includegraphics[width=\columnwidth]{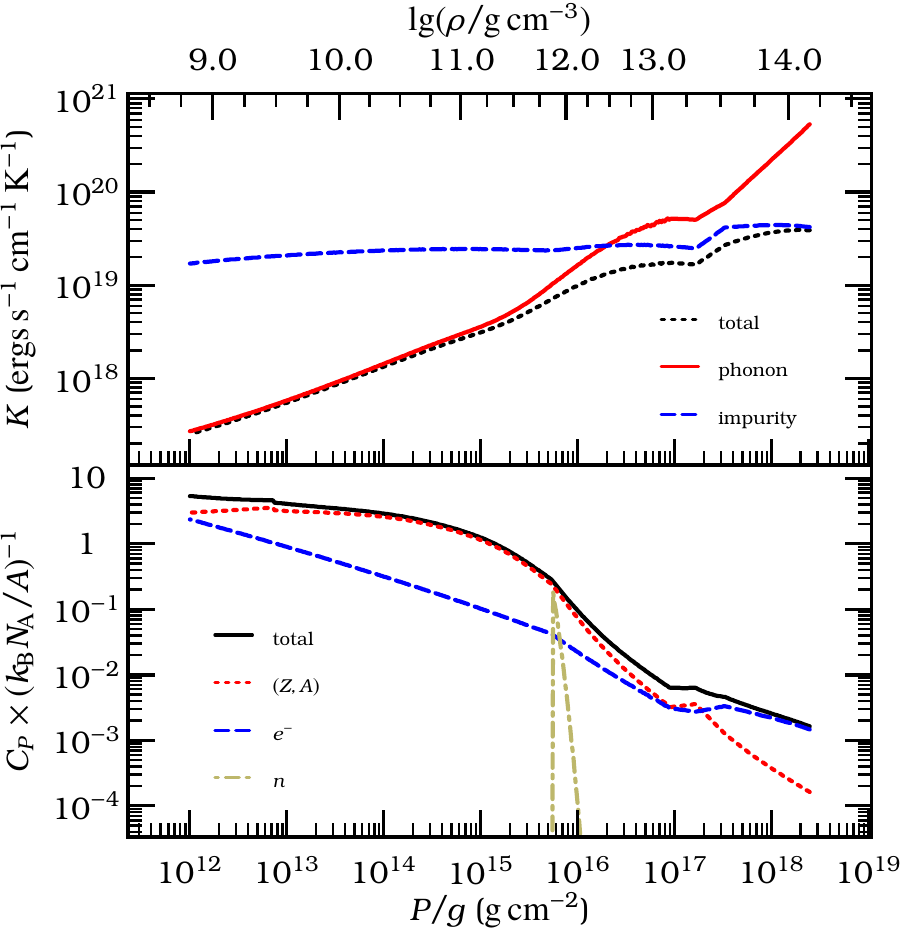}
\caption{\label{f.k-cp} Physical quantities setting the thermal diffusion time (cf.\ eq.\ [\ref{e.analytical-tau-outer}]). The top panel shows the total thermal conductivity (\emph{dotted line}), and the conductivity resulting from electron-phonon (\emph{solid line}) and electron-impurity (\emph{dashed line}) scattering.  These quantities are computed for the temperature at the end of the outburst for the run corresponding to \mx. The bottom panel displays the specific heat (\emph{solid line}), in units of $\kB \NA/A$, and the contributions from ions (\emph{dotted line}), electrons (\emph{dashed line}), and neutrons (\emph{dot-dashed line}).  The slight step in $C_{P}$ (\emph{bottom panel}) at $P/g \lesssim 10^{13}\nsp\columnunit$ is where the ions crystallize. The top axis of the plot indicates the density as a function of $P/g$. The scale is not linear because of the change in the effective polytropic index in the inner crust.}
\end{figure}

The origin of the broken power law nature of the lightcurve lies in the change in slope of the thermal time with depth that occurs close to neutron drip (see the lower panel of Fig.~\ref{f.T-tau}; neutron drip occurs at $P/g\approx 5\times 10^{15}\ {\rm g\ cm^{-2}}$). The decrease in slope is primarily due to the suppression of $C_{p}$ in the inner crust, shown in Figure~\ref{f.k-cp}. The ion contribution to the specific heat (\emph{bottom panel, dotted line}) decreases on going to higher densities roughly as $(T/\Theta_{D})^{3}$, where the Debye temperature $\Theta_{D} \propto \Theta_p=(\hbar/\kB)\left[4\pi Z^{2}e^{2}n_{\mathrm{ion}}/(A\mb)\right]^{1/2}$, the plasma temperature of the ions. We assume that the neutrons in  the inner crust are superfluid, in which case they have a negligible contribution to the heat capacity (Fig.~\ref{f.k-cp}, \emph{bottom panel, dot-dashed line}). The thermal time also depends on the thermal conductivity, which changes from being set by phonon scattering in the outer crust to impurity scattering in the inner crust (\emph{top panel, dashed line}). Electron-electron scattering, although included in our calculations, is not a significant component of the total thermal conductivity \citep{Shternin2007Neutron-star-co}, and we do not show it in Fig.~\ref{f.k-cp}. The slight step in the ion specific heat at $P/g \lesssim 10^{13}\nsp\columnunit$ (Fig.~\ref{f.k-cp}, \emph{bottom panel}) is caused by the liquid-solid transition in the crust.  Our code does not follow the crystallization front and hence does not include the latent heat. The depth where crystallization occurs is so shallow, however, that this omission does not appreciably affect the lightcurve, unlike the case for the cooling of white dwarf stars \citep[see][and references therein]{Chabrier1998Review-on-White,Hansen2004The-astrophysic}. We note that observations taken shortly ($\lesssim 10\nsp\unitday$) after the end of the outburst could potentially detect the effect of the latent heat; this would provide an independent constraint on the temperature in the crust and a check on the value of the plasma parameter $\Gamma$ at which the ions crystallize (see Appendix~\ref{s.eos}).

With some approximations, the same arguments allow us to make an analytic approximation to the lightcurve. 
The slope of the cooling curve can be written
\begin{equation}\label{e.slope}
\frac{\dif\ln\Teffinf}{\dif\ln t} =\left(\frac{\dif\ln\Teffinf}{\dif \ln T}\right)\left(\frac{\dif\ln T}{\dif\ln y}\right)\left(\frac{\dif\ln y}{\dif\ln \tau}\right). 
\end{equation}
The first factor on the right hand side is the slope of the $T_\mathrm{eff}$--$T$ relation, $\dif\ln\Teffinf/\dif\ln T\approx 0.45\textrm{--}0.63$ (Fig.~\ref{f.teff-tb}). The second factor is the temperature gradient in the initial model. The third factor is the dependence of thermal time with column depth. We can obtain this analytically by noting that during the early part of the lightcurve, when the cooling wave is in the outer crust, we can approximate the heat capacity as $C_{P} \approx 3\kB/(A\mb)$ the classical heat capacity of a lattice, and use an approximate expression for the phonon conductivity (see eq.~[\ref{e.Wiedemann}]--[\ref{e.phonon-freq}]). Inserting these expressions into the expression for $\tau$ (eq.~[\ref{e.tau}]), we find\footnote{In principle, we could include the dependence of $\Ye$ on $y$ (eq.~[\ref{e.Ye}]), but for simplicity we shall leave it free in this formula. This expression agrees well with our numerical calculations (Fig.~\ref{f.T-tau}, \emph{bottom panel}).}
\begin{equation}\label{e.analytical-tau-outer}
\tau \approx 34\nsp\unitday \cdot y_{14}^{3/4} \left(\frac{2.3}{g_{14}}\right)^{5/4} \left(\frac{\Ye}{0.5}\right) \left(\frac{60}{A}\right),
\end{equation}
where $g_{14} = g/10^{14}\nsp\cm\usp\second^{-2}$, giving $d\ln \tau/d\ln y=3/4$.

A measurement of the slope during the early part of the lightcurve directly measures, therefore, the temperature gradient in the outer crust at the end of the outburst, 
\begin{equation}\label{e.tempgrad}
{\dif \ln T\over \dif\ln y}\approx {15\over 11}{\dif\ln T^\infty_\mathrm{eff}\over \dif\ln t}.
\end{equation}
The fact that the effective temperature is decreasing with time, implies that temperature decreases with depth in the crust: the temperature profile in the outer crust is inverted during outburst. 
If the temperature increased with depth in the crust, equation (\ref{e.tempgrad}) implies that we would see an increasing temperature with time, and this can in fact be seen in the numerical models of  \citet[][see their Fig.~3]{rutledge.ea.01:ks1731}.

A different way to write equation (\ref{e.tempgrad}) is in terms of the inwards flux in the outer crust, since the flux determines the temperature gradient. We write the flux as $F = -K\dif T/\dif r$ and use the expression for $K$, eqs.~(\ref{e.Wiedemann})--(\ref{e.phonon-freq}) to obtain
\begin{equation}\label{e.analytical-flux-outer}
\frac{\dif\ln T}{\dif\ln y} = 0.04 \left(\frac{F}{10^{21}\nsp\ergspersecond\usp\cm^{-2}}\right)
\left(\frac{\Ye}{0.5}\right)\left(\frac{2.3}{g_{14}}\right)^{5/4} y_{14}^{-1/4} T_{8}^{-1}.
\end{equation}
giving
\begin{equation}\label{e.slope-Teff}
\frac{\dif\ln\Teffinf}{\dif\ln t} = 0.03 \left(\frac{F}{10^{21}\nsp\ergspersecond\usp\cm^{-2}}\right),
\end{equation}
where we have suppressed the other factors from equation~(\ref{e.analytical-flux-outer}) and we use $\dif\ln\Teffinf/\dif\ln T = 0.55$. As an example, we plot a line representing this slope in Fig.~\ref{f.mxb1659-model} (\emph{grey dashed line}), but using $\dif\ln\Teffinf/\dif\ln T = 0.45$, as appropriate at this \Teff. The slope of the early-time power law directly measures the flux in the  rust.

\begin{figure}[htbp]
\includegraphics[width=\columnwidth]{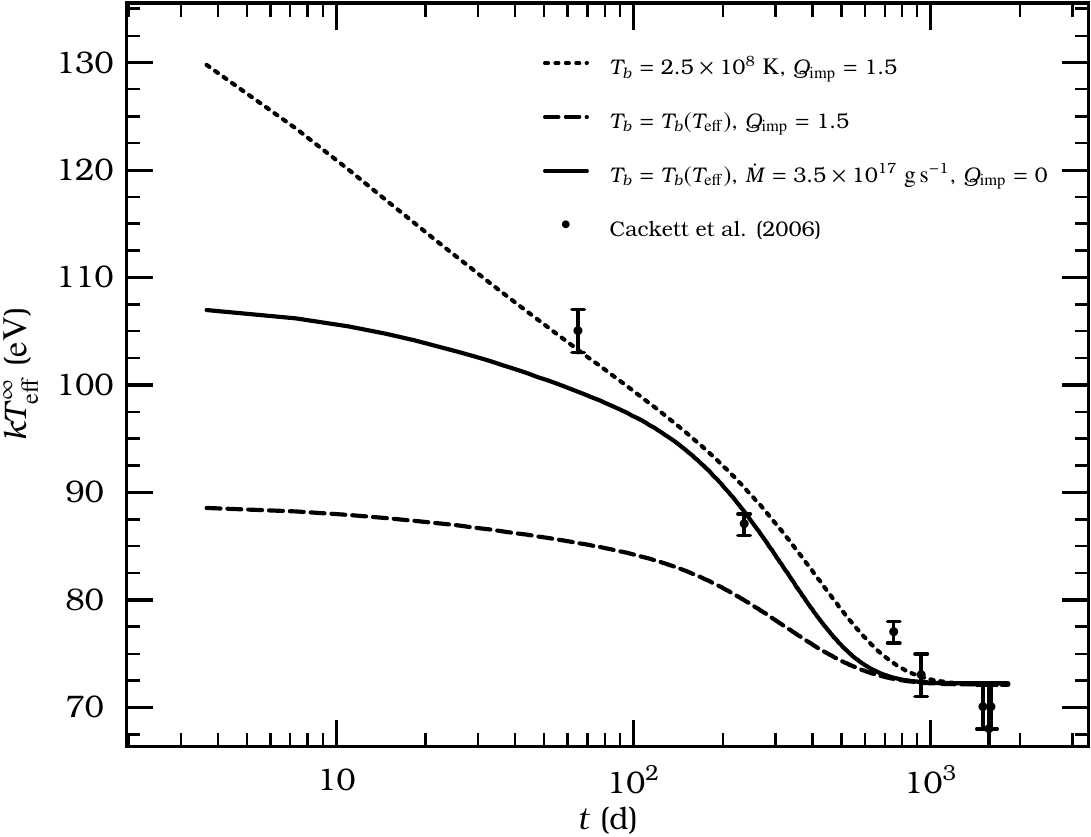}
\caption{\label{f.ks-compare-lightcurve}Our fit to the lightcurve for \ks\ (\emph{dotted line}) compared to calculations for which the outer boundary condition during outburst was replaced with the same $T_{b}(\Teff)$ relation used during quiescence (\emph{dashed, solid lines}). For the case where other parameters were held fixed (\emph{dashed line}), the lightcurve falls markedly below the observed values.  When the outburst accretion rate is increased to $3.5\ee{17}\nsp\grampersecond$ and the impurity parameter decreased to $\Qimp=0$ (\emph{solid line}) a better agreement is found with the observed lightcurve.  A clear distinction between these scenarios could be observed during the first fortnight of quiescence.}
\end{figure}

Our inference that the temperature gradient in the crust is inverted, i.~e., the temperature decreases with depth, also implies that an inward-directed heat flux enters the crust from the top.  To show how this is critical for the early decay, Figure~\ref{f.ks-compare-lightcurve} compares our original fit (\emph{dotted line}) for \ks\ with runs in which we do not hold the temperature at the top of the crust fixed, but rather use the same boundary condition (see Fig.~\ref{f.teff-tb}) during both outburst and quiescence (\emph{dashed and solid lines}). We first changed only the condition on $T_{b}$, while holding all other parameters fixed (\emph{dashed line}). Although the break in the lightcurve occurs on the same timescale, the calculation with deep heating only is too faint to match observations. We then increased the accretion rate to $\Mdot = 3.5\ee{17}\nsp\grampersecond$, which gives a crust heating rate comparable to that used by \citet{Shternin2007Neutron-star-co}. Because the higher temperature decreases the thermal conductivity, we set $\Qimp = 0$ to compensate. The resulting solution  (\emph{solid line}) gives a better fit, although its early time behavior still undershoots the first observation. 
For both latter cases (\emph{dashed and solid lines}), the temperature profile is inverted for $P/g \gtrsim 10^{13}\nsp\columnunit$; that is, most of the heat produced in the outer crust during outburst is conducted inward. As a result, $\Teffinf$ is decreasing with time starting a few days after the end of outburst in these cases.

\begin{figure}[htbp]
\includegraphics[width=\columnwidth]{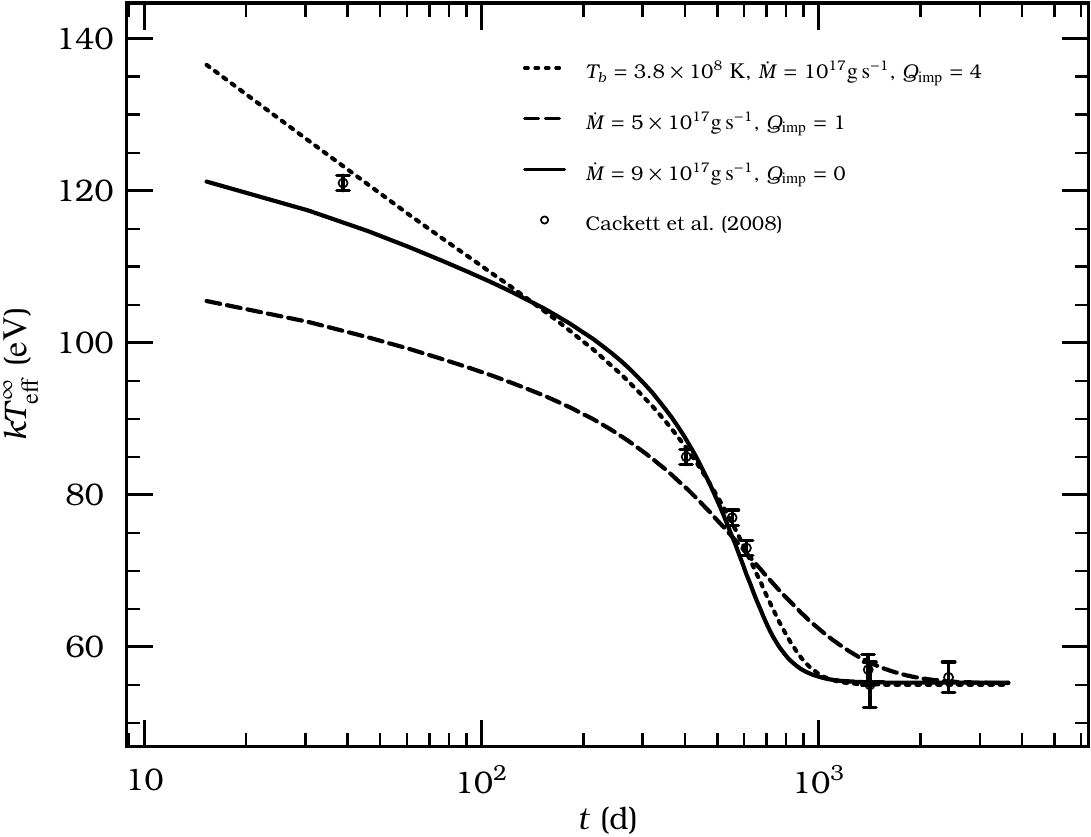}
\caption{\label{f.mxb-compare-heating}A comparison between the observed lightcurve of \mx\ and several trial numerical solutions: 1) our best-fit solution (\emph{dotted line}; cf.~Fig.~\protect\ref{f.mxb1659-model}) with $T_{b}=3.8\ee{8}\nsp\K$ and $\Mdot=10^{17}\nsp\grampersecond$; a solution with $\Mdot=5\ee{17}\nsp\grampersecond$, $\Qimp=1$, and $T_{b}=T_{b}(\Teff)$ (\emph{dashed line}); and a solution with $\Mdot=9\ee{17}\nsp\grampersecond$, $\Qimp=0$, and $T_{b}=T_{b}(\Teff)$.}
\end{figure}

As noted by \citet{Shternin2007Neutron-star-co}, the observed lightcurve of \ks\ can be explained without recourse to an inward-directed heat flux provided that $\Mdot$ is larger than the value we assume in this paper (cf.\ Fig.~\ref{f.ks-compare-lightcurve}). We have made the same comparison for \mx\ (Fig.~\ref{f.mxb-compare-heating}): we used the same $T_{b}(\Teff)$ relation during both outburst and quiescence, and then increased $\Mdot$ while decreasing $\Qimp$. In addition to our best-fit solution with fixed $T_{b}$ (\emph{dotted line}), we show the case for $\Mdot=5\ee{17}\nsp\grampersecond$ and $\Qimp=1$ (\emph{dashed line}) and $\Mdot=9\ee{17}\nsp\grampersecond$ and $\Qimp=0$. Without a fixed $T_{b}$ during outburst (or equivalently a sizeable inward-directed heat flux in the shallow outer crust), we do not find an acceptable numerical solution consistent with the observed \mx\ lightcurve.

To illustrate this point, we computed, using equation~(\ref{e.slope-Teff}), the required flux necessary to match the observed power-law slope between the first two observations.  For \ks\ the first two observations imply a flux of $0.7\nsp\MeV/\mb$ for $\Mdot = 10^{17}\nsp\grampersecond$. For \mx, the flux required to match the observed power-law slope is $0.8\nsp\MeV/\mb$ for $\Mdot = 10^{17}\nsp\grampersecond$. Our numerical solutions are consistent with this estimate: our numerical model of the \ks\ lightcurve (Fig.~\ref{f.ks-lightcurve}) has a local inward flux of $0.5\nsp\MeV/\mb$, while our model of the \mx\ lightcurve (Fig.~\ref{f.mxb1659-model}) has $1.1\nsp\MeV/\mb$.  This value of the flux is well above the estimates available from electron captures \citep{Gupta2006Heating-in-the-}. Moreover, this flux would have to emanate from a depth for which the thermal time is less than the time of the first observation, which is $P/g \lesssim 2\ee{14}\nsp\columnunit$, corresponding to $\rho \lesssim 3\ee{10}\nsp\grampercc$. This is a lower density than that of many crust electron captures \citep{Gupta2006Heating-in-the-,Haensel2008Models-of-crust} and even that of light element pycnonuclear reactions, such as \oxygen[24] \citep{Horowitz2007Fusion-of-neutr}. Hence, although a larger \Mdot\ reduces the total amount of heating per nucleon required, matching the first data point is still difficult. Observations of both sources are too sparse on timescales $\lesssim 10^{2}\nsp\unitday$, however, to draw firmer conclusions about the existence and nature of this heating. Observations occurring within the first two weeks after the outburst ends are critical for constraining the depth and strength of heat sources in the outer crust.

\section{Constraints on the model parameters}\label{s.constraints}

In Figures \ref{f.mxb1659-model} and \ref{f.ks-lightcurve}, we show models that fit the data well for \mx\ and \ks. We now address the uncertainties in the fitted model parameters. The cooling models have a range of physics input. To allow us to investigate the full range of parameter space, we adopt the simplified model of the cooling curves described in \S 2.4 as it allows us to rapidly generate a lightcurve without running the full time-dependent simulation. 

To calculate the constraints on model parameters, we adopt a Markov Chain Monte Carlo (MCMC) method, using the Metropolis-Hastings algorithm to generate a sequence of samples from the posterior probability distributions \citep[e.~g.,][]{Gregory2005A-Bayesian-Anal}. 
Although we have a small number of parameters which would allow a grid search of the chi-squared space, we prefer the MCMC method for its simplicity in implementation and generating marginalized probability distributions. We run the chains multiple times from different starting points to check the convergence and robustness of the resulting probability distributions. Typically, we find that $\sim 10^4$ samples are adequate, although we have run the chains longer to check convergence.

Although the simplified model lightcurves agree well with the numerical results (as shown for a specific example in Figure \ref{f.mxb1659-model}), the agreement is not exact, and as a result the best-fit parameters derived from the two models are different. For example, whereas the best fit lightcurve, for \mx, from the time-dependent simulation has $\Qimp=4$ (Fig.~\ref{f.mxb1659-model}), the simplified model gives a best fit value $\Qimp=2$. In this section our focus is, however, on how well we can constrain the main parameters of the model, and how each of these parameters changes the lightcurve, rather than the best-fit values. By running a sample of comparison models, we have checked that the uncertainties in the parameters derived from the simplified model and time-dependent simulations are similar.

\subsection{Constraints for a given hydrostatic structure}\label{s.constraints-fixed-hydro}

For a given hydrostatic structure, which is set by the neutron star mass and radius, the three main parameters that affect the lightcurves are the temperatures at the top of the crust $T_b$, the core temperature $T_c$, and the impurity parameter $\Qimp$. We assume for now that the accretion rate, which determines the overall heating rate in the crust (see Appendix), is determined from the observed X-ray luminosity. We fix the neutron star mass $M=1.6\ M_\odot$ and radius $R=11.2\ {\rm km}$, and the outburst accretion rate $\dot M=10^{17}\ {\rm g\ s^{-1}}$ for both sources. These values match the parameters used in the time-dependent code to calculate the models shown in Figures  \ref{f.mxb1659-model} and \ref{f.ks-lightcurve}. The corresponding redshift factor is $1+z=1.32$ and gravity is $g_{14}=2.3$.

\begin{figure}[htbp]
\includegraphics[width=\columnwidth]{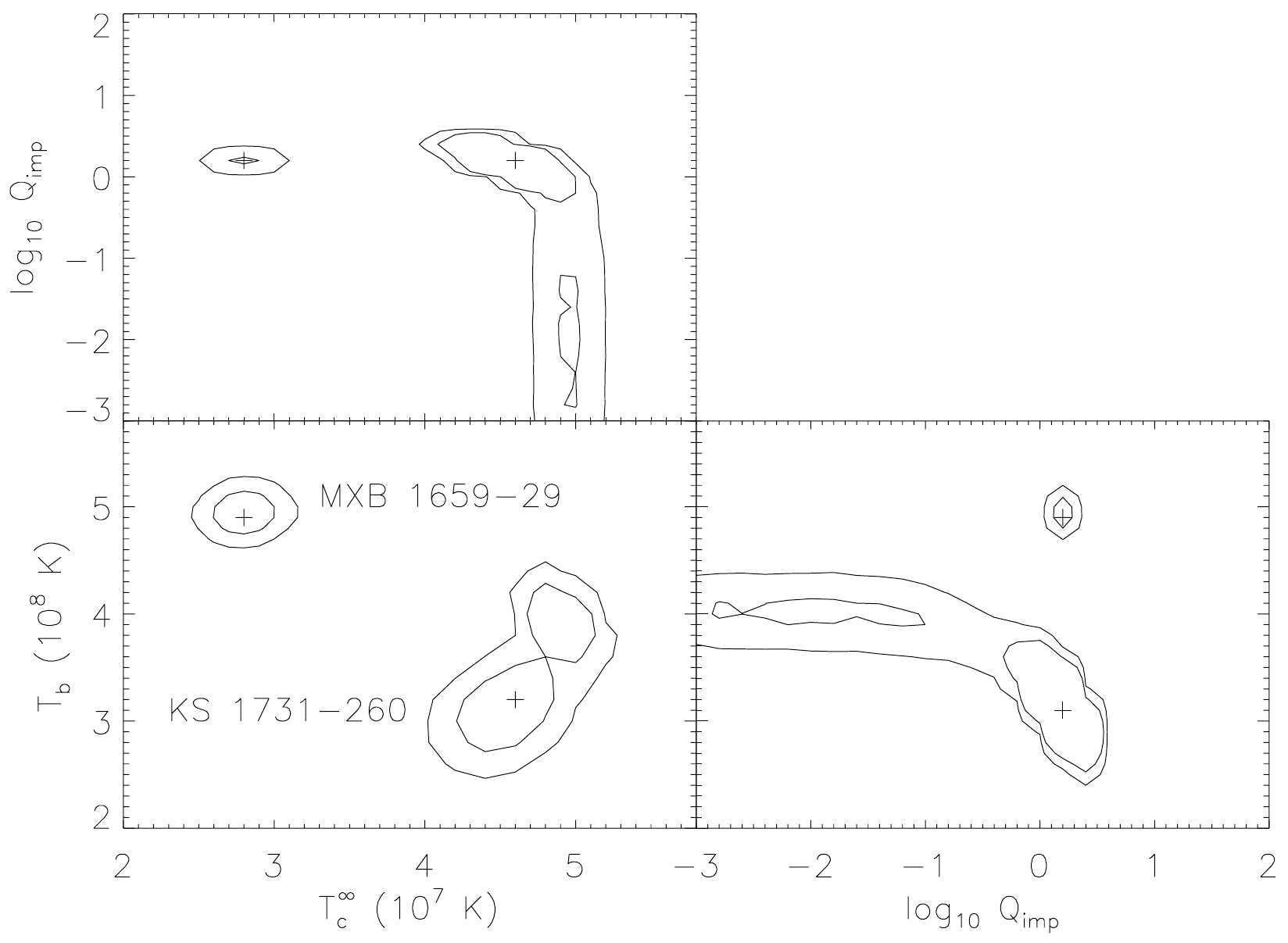}
\caption{Probability densities for $\Qimp$, $T_c^\infty$ and $T_b$ for MXB~1659-29 and KS~1731-260. The peak of the probability distribution is marked by a cross, and the contours show the 68\% and 95\% confidence intervals. The prior probability for $\Qimp$ is taken to be uniform in log from $\Qimp=10^{-3}$ to $10^2$. For each source, we fix the outburst accretion rate at $\dot M=10^{17}\ {\rm g\ s^{-1}}$ and the neutron star mass and radius at $M=1.6\ M_\odot$ and $R=11.2\ {\rm km}$.
\label{f.qmap}}
\end{figure}

Figure \ref{f.qmap} shows the resulting probability distributions for $\Qimp$, $T_c^\infty$ and $T_b$. The values of $T_c^\infty$ and $T_b$ are well-constrained in each case. These two temperatures are constrained by the need to match the first and last data points in each set of observations. The first data point gives a snapshot of the temperature near the top of the crust (thermal time of tens of days), whereas the last point gives a measurement of the core temperature (we find that the lightcurve has leveled off in both sources in our models).

\begin{figure}[htbp]
\includegraphics[width=\columnwidth]{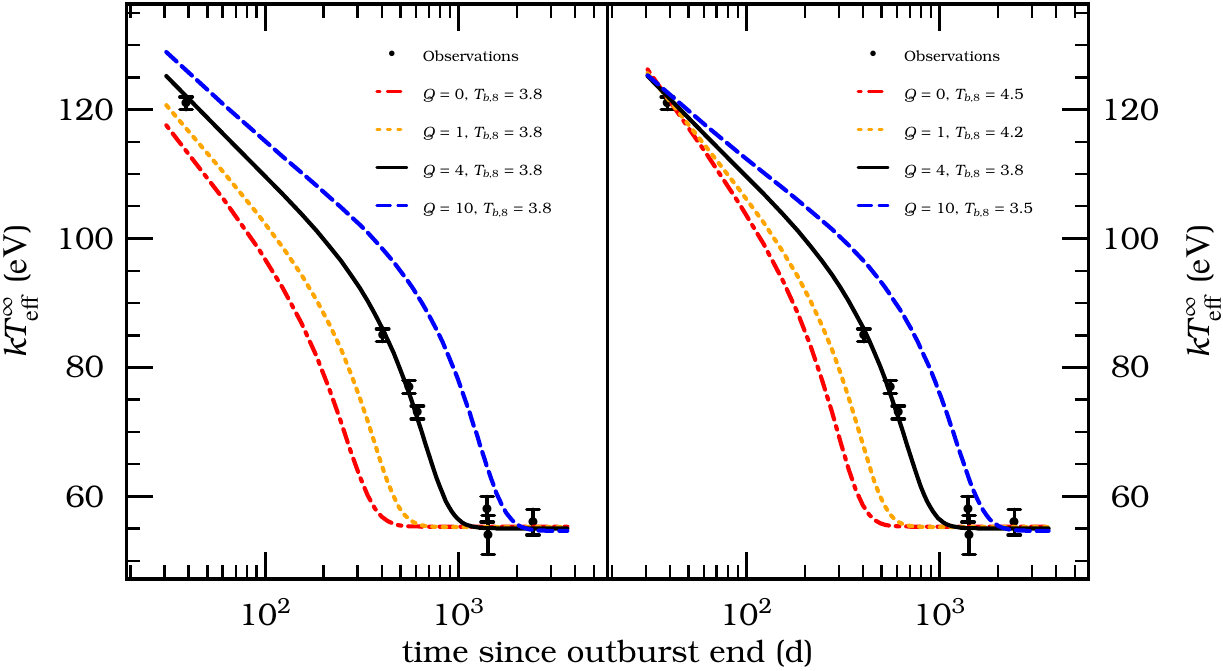}
\caption{Lightcurves of \mx, for different choices of the impurity parameter $\Qimp$ in the crust.  In both case, we show the best fit model (\emph{solid line}).  The other solutions have $\Qimp=0$ (\emph{dot-dashed line}), 1 (\emph{dotted line}), and 10 (\emph{dashed line}).  The \emph{left-hand panel} shows the case for which all other parameters are held constant; in the \emph{right-hand panel} the temperature at a column $\ytop=10^{12}\nsp\grampercc$ was adjusted so that all solutions matched the first data point.
\label{f.mxb-varQ}}
\end{figure}

The impurity parameter $\Qimp$ is very tightly constrained in \mx. To understand why this is the case, we show in Figure \ref{f.mxb-varQ} several lightcurves from our time-dependent simulations for \mx\ with different values of \Qimp. In the left panel, we keep all other parameters fixed as we vary \Qimp; in the right panel, we adjust $T_b$ to match the first data point as we vary \Qimp. The effect of increasing \Qimp\ is to delay the cooling. This can be understood in terms of the initial temperature profile at the end of the outburst. When the crust has a larger impurity level, the inner crust must be hotter to be able to conduct the heat released in the crust into the core. This increase in the inner crust temperature leads to a hotter outer crust, with a shallower temperature gradient, giving a lightcurve that falls less quickly. 

\begin{figure}[htbp]
\includegraphics[width=\columnwidth]{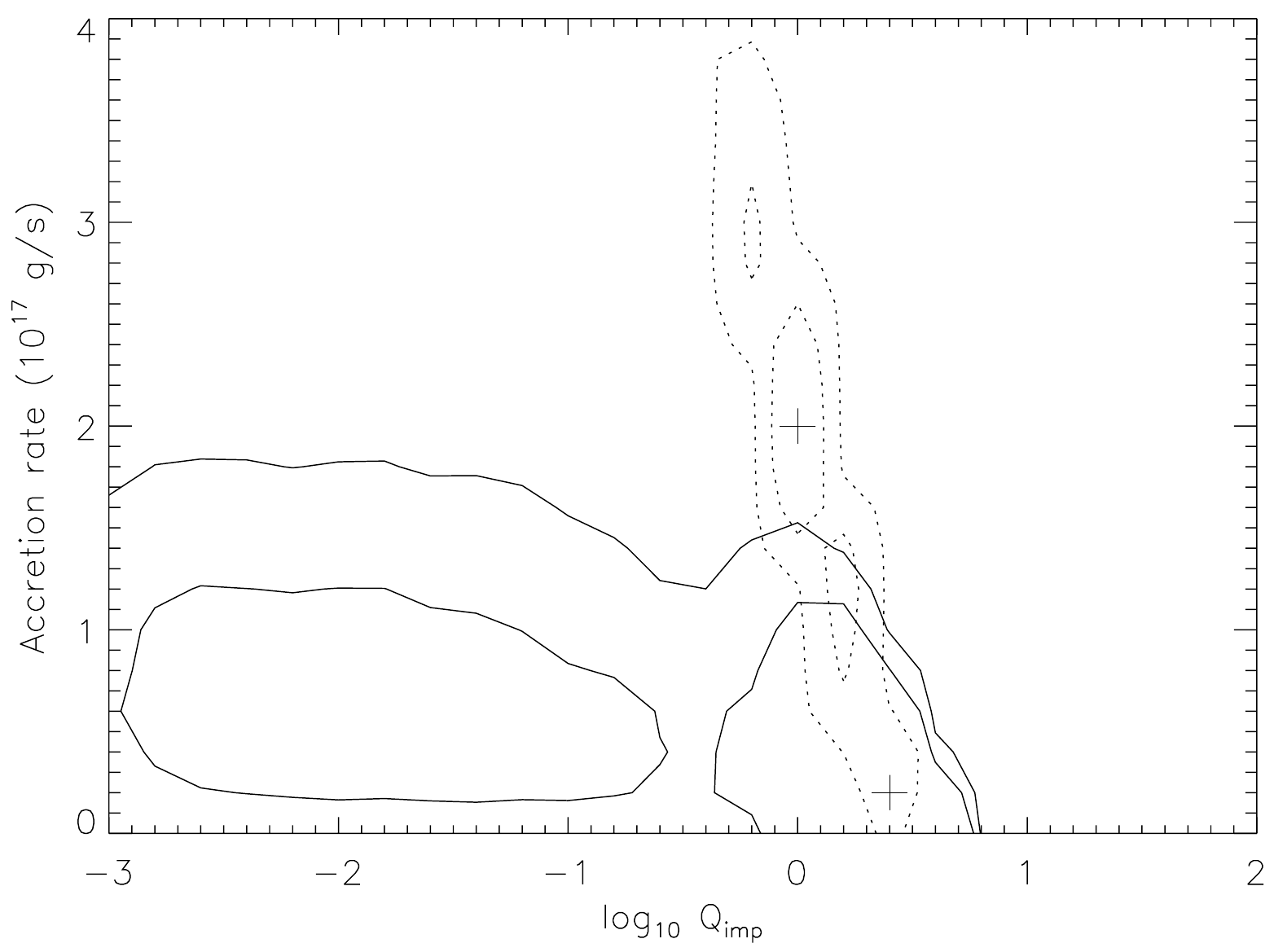}
\caption{The joint probability distribution of $\Qimp$ and $\dot M$ for \mx\ (dotted contours) and \ks\ (solid contours). In each case, the peak of the probability distribution is indicated by a cross; the two contours enclose 68\% and 95\% of the probability. 
\label{f.qmap3}}
\end{figure}

As Figure \ref{f.qmap} shows, the correlations between the fitted parameters are such that larger \Qimp\ values lead to lower values of $T_c$ and $T_b$, i.e. to compensate for the delayed cooling due to increase in \Qimp, the overall temperature scale set by $T_c$ and $T_b$ decreases. In \ks, the probability distribution of \Qimp\ has a peak at a similar value to \mx, but with a long tail to small values of \Qimp. In fact, as can be seen in Figure \ref{f.qmap3}, the fits are not sensitive to the impurity parameter for \Qimp$\lesssim1$, which results in a flat probability distribution in $\log$ \Qimp, reflecting the assumed prior. For both sources, \Qimp\ values larger than 10 are ruled out. 

For \mx, we have used the temperatures derived by Cackett et al.~(2008) assuming a distance to the source of $10\ {\rm kpc}$. In that paper, spectral models for different distances $d=5$ and $13\ {\rm kpc}$ are considered, which leads to a systematic decrease or increase in the effective temperatures by $10$--$20$\%. The reason that the fitted effective temperatures depend on distance is that the peak of the thermal spectrum lies outside the X-ray band, making the fitted temperature sensitive to the overall luminosity scale. To investigate the effect of such systematic variations, we have calculated the constraints on the models with the effective temperatures for \mx\ all decreased or increased by 20\%. The effect is to change the central value of each distribution by up to 50\%, with the width staying about the same. The conclusion that \Qimp\ is of order unity is unaffected by these systematic variations.

\subsection{The accretion rate or overall heating rate in the crust}\label{s.constraints-mdot}

The accretion rate $\dot  M$ sets the overall amount of heating in the crust during the outburst. There are uncertainties in deriving $\dot M$ from the observed X-ray luminosity, and in addition, the amount of heating in the crust may differ from the $1.7\nsp\MeV$ per nucleon that we assume in our calculation (see Appendix for details). The calculations so far have taken a fixed accretion rate $\dot M=10^{17}\ {\rm g\ s^{-1}}$. Instead, we now calculate the constraints on $\dot M$ assuming a uniform prior probability for $\dot M$ between $0$ (i.e.~no deep heating) and $10^{18}\nsp\grampersecond$ (ten times our fiducial rate). The results are shown in Figure \ref{f.qmap3}, in which we give the derived joint probability distribution for $\dot M$ and \Qimp\ for each source. The temperatures $T_b$ and $T_c$ are not sensitive to variations in $\dot M$, since they are essentially fixed by the first and last observed values of $T^\infty_{\rm eff}$.

For both sources, we find an anti-correlation between $\dot M$ and \Qimp\ in the best-fitting solutions. The explanation for the anti-correlation is that an increased $\dot M$ gives an increased heating rate, making the inner crust hotter. To compensate for this, \Qimp\ must decrease, cooling the inner crust by making it easier for heat to be conducted into the core. 

The values of $\dot M$ derived from the cooling curves match well with the accretion rates derived from observations of the persistent X-ray luminosity during outburst. For \mx, the range of flux observed during the outburst was $\approx (0.4\textrm{--}1)\times 10^{-9}\nsp\ergspersecond\usp\cm^{-2}$ ($2.5\textrm{--}25\nsp\keV$) \citep{Galloway2008Thermonuclear-t}. \citet{Galloway2008Thermonuclear-t} found a distance of $12\pm 3$ kpc for this source, assuming that the peak flux of photospheric radius expansion bursts corresponds to the pure helium Eddington luminosity. Taking this distance and assuming a bolometric correction of a factor of 2, typical for these sources, gives $\dot M\approx (0.7$--$1.8)\times 10^{17}\ {\rm erg\ s^{-1}}$. The agreement with the constraints from the cooling curve is good, although lower than the maximum of the probability distribution for $\dot M$.
 
For \ks, \citet{Galloway2008Thermonuclear-t} give a range of bolometric flux $1.6\textrm{--}10\times 10^{-9}\ {\rm erg\ cm^{-2}\ s^{-1}}$, which for their distance $7.2\pm 1\ {\rm kpc}$ gives a range of accretion rates during outburst of $0.5\textrm{--}3\times 10^{17}\ {\rm g\ s^{-1}}$. A separate check on this value is that at a flux level of $2.1\times 10^{-9}\ {\rm erg\ cm^{-2}\ s^{-1}}$, a very regular sequence of X-ray bursts was seen, similar to the source GS~1826-24, which is known for being a very regular burster. Assuming an ignition mass of $10^{21}\ {\rm g}$ for these regular bursts, which had a recurrence time of $2.59\pm 0.06\ {\rm h}$, we find $\dot M=1.1\times 10^{17}\ {\rm g\ s^{-1}}$, consistent with the X-ray flux. During the final $\gtrsim 1$ year of the outburst, the flux was in the lower end of the flux range quoted earlier, so we expect the relevant value for the crust heating at the end of the outburst to be $\lesssim 10^{17}\ {\rm g\ s^{-1}}$, in good agreement with Figure \ref{f.qmap3}.

An interesting aspect of our results is that both sources allow solutions with low accretion rates much smaller than the accretion rates derived from the X-ray observations. Assuming that the observed accretion rate is within a factor of two of the true accretion rate onto the neutron star, this means that both cooling curves are consistent with a much lower amount of deep crustal heating than assumed in our models. In these models, however, a lower level of deep crustal heating from reactions in the crust is compensated by a larger inward heat flux from the neutron star ocean, because $T_{b}$ is held fixed. In reality, the physics of the implied, unspecified heat source in the neutron star ocean that supplies this flux also depends on the accretion rate.

\begin{figure}[htbp]
\includegraphics[width=\columnwidth]{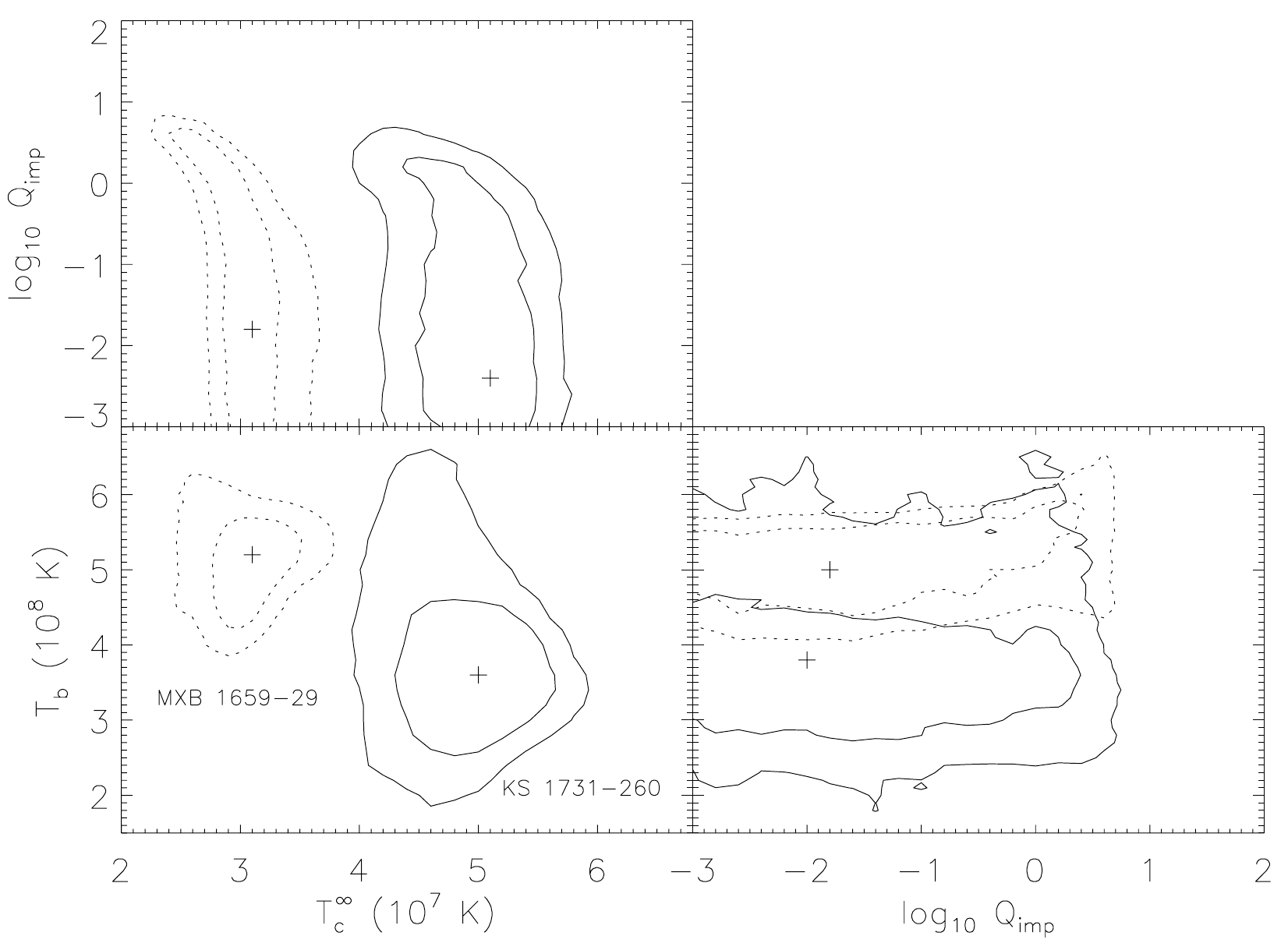}
\caption{Constraints on $T_c^\infty$, $T_b$, and \Qimp, as in Figure \ref{f.qmap}, but now including variations in $M$, $R$, and $\dot M$.
\label{f.qmap_2}}
\end{figure}

\subsection{Effect of varying neutron star mass and radius}\label{s.constraints-ns}

We now include the uncertainty in all six parameters of the model. The prior probabilities for $T_c^\infty$, $T_b$, \Qimp, and $\dot M$ are as described previously. In addition, we assume a uniform prior in the range $8$--$16\nsp\km$ for $R$ and $1.1$ to $2.5\nsp\Msun$ for $M$. The fit for \Teffinf\ used a fixed $M=1.4\nsp\Msun$ and $R=10\nsp\km$ \citep{Cackett2006Cooling-of-the-,Cackett2008Cooling-of-the-} because the quality of the data is insufficient to constrain these parameters independently; therefore, the constraints we derive should be considered indicative. The resulting constraints on $T_c^\infty$, $T_b$, and \Qimp\ are shown in Figure \ref{f.qmap_2}. This Figure should be compared with Figure \ref{f.qmap}, for which $M$, $R$, and $\dot M$ were assumed to be known. Including the uncertainty in the extra parameters broadens the probability distributions of $T_c^\infty$, $T_b$ and \Qimp. The largest effect is on the probability distribution of \Qimp. For example, the constraint on \Qimp\ is considerably relaxed for \mx. For both sources, however, \Qimp\ values greater than several are ruled out even with the additional parameters included. The central values of $T_c^\infty$ and $T_b$ are similar to the values previously derived.

\begin{figure}[htbp]
\includegraphics[width=\columnwidth]{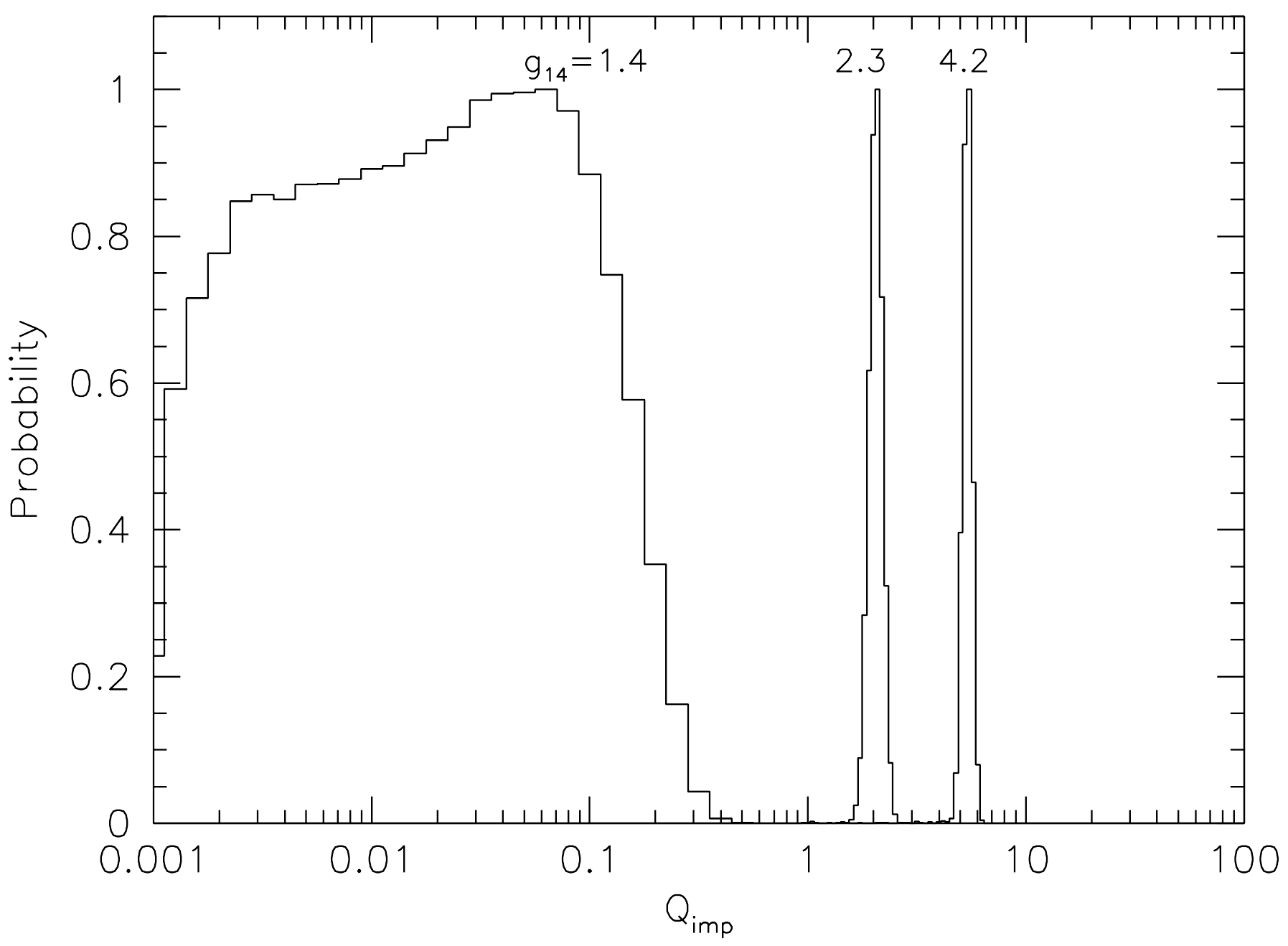}
\caption{The probability distribution of $\Qimp$ derived for \mx, for three different choices of neutron star mass and radii. Left to right, in order of increasing surface gravity, they are (i) $M=1.4\ M_\odot$, $R=13\ {\rm km}$, $g_{14}=1.4$, $1+z=1.21$ (ii) $M=1.6\ M_\odot$, $R=11.2\ {\rm km}$, $g_{14}=2.3$, $1+z=1.32$ and (iii) $M=2\ M_\odot$, $R=10\ {\rm km}$, $g_{14}=4.2$, $1+z=1.57$. In each case, the accretion rate is fixed at our fiducial value $\dot M=10^{17}\ {\rm g\ s^{-1}}$. Note that the spectral fits used to obtain \Teffinf\ assume a fixed value of $M = 1.4\nsp\Msun$ and $R=10\nsp\km$.
\label{f.qmapfig_mr}}
\end{figure}

The sensitivity of the derived value of \Qimp\ on $M$ and $R$ is illustrated for \mx\ in Figure \ref{f.qmapfig_mr} (we see the same effect for the \ks\ data). We show the derived probability distribution for \Qimp\ for three different choices of neutron star mass and radius. In each case, we keep the accretion rate fixed at our fiducial value $\dot M=10^{17}\ {\rm g\ s^{-1}}$. The allowed values of \Qimp\ increase with increasing surface gravity. This can be understood by considering the thermal time from a given density to the surface, which depends on the thickness of the layer and therefore varies with surface gravity \citep{Lattimer1994Rapid-cooling-a}. Rewriting the integral for the thermal time, equation~(\ref{e.analytical-tau-outer}), as an integral over pressure gives $\tau^\infty\propto (1+z)/g^2\propto R^4M^{-2}(1+z)^{-1}$. An increase in surface gravity shortens the cooling time, and \Qimp\ must increase to bring it back into agreement with the observed value.

\begin{figure}[htbp]
\includegraphics[width=\columnwidth]{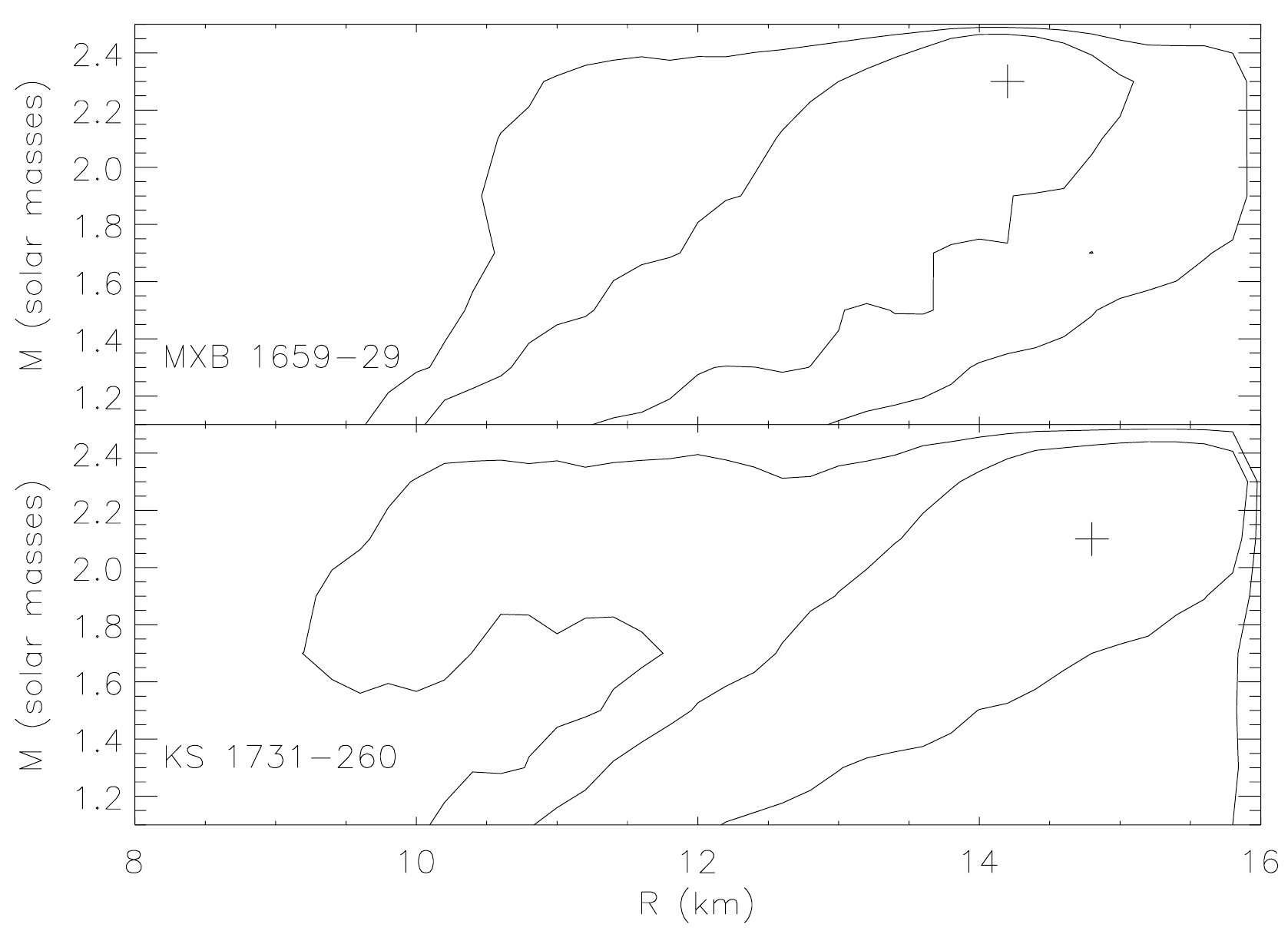}
\caption{Constraint on the neutron star mass and radius. We assume a constant prior in mass between $1.1$ and $2.5\ M_\odot$ and in radius between $8$ and $16\ {\rm km}$. The peak of the probability distribution is marked with a cross, and the contours enclose 68\% and 95\% of the probability. Note that the spectral fits used to obtain \Teffinf\ assume a fixed value of $M = 1.4\nsp\Msun$ and $R=10\nsp\km$. 
\label{f.mr}}
\end{figure}

Figure \ref{f.mr}  shows the joint probability density for $M$ and $R$ for each source. Although $M$ and $R$ are only weakly constrained, we see that the best-fitting values of $M$ and $R$ are correlated. The mass and radius enter the calculation of the lightcurve in several places besides the thermal time $\tau^\infty$. The relation between crust temperature and $T_\mathrm{eff}^\infty$ depends on the surface gravity; for a fixed crust temperature, $T_\mathrm{eff}^\infty\propto g^{1/4}/(1+z)$. The initial temperature profile also changes with gravity. Using the Newtonian equations for the steady-state thermal profile, we see that $\dif T/\dif P=(1/g)(3\kappa F/4acT^3)$, $\dif F/\dif P=-\epsilon/g$, so that the increase in flux due to the deep heating is smaller by a factor $g$, and the change in temperature for a given flux is smaller by a factor $g$. The combination of these different effects results in the observed correlation between the best fitting values of $M$ and $R$. By inspection we find that the slope of the relation is well-described by $g\propto (1+z)^3$.

\section{Discussion and Conclusions}\label{s.conclusions}

We have presented numerical simulations of the cooling of the neutron star crust in both \ks\ and \mx\ following the end of long accretion outbursts. Our main results are as follows.
\begin{enumerate}
\item The lightcurve of a cooling crust is a broken power-law going to a constant at late times. The luminosity at late times is set by the neutron star core temperature. The slope of the early part of the lightcurve provides a direct measure of the flux in the outer crust during outburst (eq.~[\ref{e.slope-Teff}]). The time of the break is set by the transition from a classical to quantum crystal, close to neutron drip. The good fit of our models to the data suggests that the neutrons in the inner crust do not contribute significantly to the heat capacity, as expected if they were superfluid. Observations of cooling quiescent neutron stars thus provide new evidence for the existence of a neutron superfluid throughout the inner crust.

\item As our models show, the observations to date probe the thermal relaxation timescale of the inner crust. The cooling timescale increases with increasing \Qimp, potentially giving a tight constraint on this parameter. The fits to the lightcurves of \mx\ and \ks\ both require $\Qimp<10$, in agreement with the result of \citet{Shternin2007Neutron-star-co} for \ks. For our fiducial model, which has neutron star parameters $M=1.6\nsp\Msun$, $R=11.2\nsp\km$, and outburst accretion rate $\Mdot =10^{17}\nsp\grampersecond$, the best fit values are $\Qimp=4$ for \mx, and $\Qimp=1.5$ for \ks. Reducing the surface gravity or increasing the accretion rate allows smaller values of \Qimp. Impurity scattering sets the thermal conductivity of the inner crust, and so our measurement of \Qimp\ refers to the conductivity in the inner crust, particularly close to neutron drip where the thermal time corresponds to the time of the break in the cooling curve. Interestingly, the values of \Qimp\ derived for both \ks\ and \mx\ are very similar, and may indicate that a robust outcome of nuclear processing in an accreted crust is an inner crust impurity parameter of order unity, as suggested by calculations of nuclear transitions in the inner crust \citep{jones:end-point-rp-process,Gupta2008Neutron-Reactio}.

\item The flux required to match the power-law slope between the first and second observations is much larger, however, than expected from models of deep heating \citep{Gupta2006Heating-in-the-,Haensel2008Models-of-crust}.  For \ks, we find that the lightcurve at $\gtrsim 100\nsp\unitstyle{d}$ post-outburst can be fit using standard models of deep heating, if the accretion rate is larger than our estimate of $10^{17}\nsp\grampersecond$, in agreement with the findings of \citet{Shternin2007Neutron-star-co}. We do not find such a solution for \mx, however; indeed setting the outer boundary condition to $T_{b}=T_{b}(\Teff)$ drives the outburst accretion rate to roughly the Eddington limit.  To obtain an adequate fit to the data, we require the temperature in the outer crust to be decreasing inwards, implying that an inward-directed heat flux enters the crust from the top. Moreover, our Markov Chain Monte Carlo fits with our simplified model (\S~\ref{s.constraints}) find best fit solutions with rather large values of $T_{b}$, so that the temperature profile is inverted. Our interpretation therefore differs slightly from that of \citet{Shternin2007Neutron-star-co}. This interpretation depends, however, on the first data points in each cooling curve, and so could be relaxed if these data points are contaminated, by residual accretion for example. Our calculations show that observations taken within the first two weeks following extended outbursts are ideal for mapping out the nuclear heating in the outer crust.  It is this shallow heating that is most critical for determining the ignition depth of superbursts \citep{Gupta2006Heating-in-the-}. We shall investigate the heating required to maintain the inferred high $T_{b}$, along with its implications for nuclear processes in the neutron star ocean, in a follow-up paper.
\end{enumerate}

\acknowledgements

We thank Chuck Horowitz, Ed Cackett, Nathalie Degenaar, Sanjay Reddy, Andrew Steiner, Lars Bildsten, Gil Holder, Bob Rutledge, and Sanjib Gupta for helpful discussions. EFB and AC acknowledge the hospitality of the Institute for Nuclear Theory, where this work took shape, and by the support of the Joint Institute for Nuclear Astrophysics (\emph{JINA}) under NSF-PFC grant PHY~02-16783. EFB is supported by the National Aeronautics and Space Administration through Chandra Award Number TM7-8003X issued by the Chandra X-ray Observatory Center, which is operated by the Smithsonian Astrophysical Observatory for and on behalf of the National Aeronautics Space Administration under contract NAS8-03060, by NASA award NNX06AH79G, and by NASA ATFP grant NNX08AG76G. AC acknowledges support from an NSERC Discovery Grant, FQRNT, and the Canadian Institute for Advanced Research (CIfAR), and as an Alfred P.~Sloan Research Fellow.

\appendix

\section{Crust Microphysics}\label{s.crust-microphysics}

In this Appendix, we describe how we calculate the different pieces of microphysics that go into the crust models.

\subsection{Equation of state and composition profile}\label{s.eos}

We compute the equation of state for the degenerate, relativistic electrons by interpolation from tables of the free energy \citep{timmes.swesty:accuracy}. For the ions we compute the EOS using the free energy fit of \citet{chabrier98} for the liquid state and the fit of \citet{farouki93} for the crystalline state. With this choice of fits, the freezing point occurs where the free energies in the liquid and solid phase are equal, namely $\Gamma \equiv Z^2 e^{2}/(akT) = 178$,  where $a$ is the ion-sphere radius. Note that our fit to the liquid-phase free energy does not include corrections \citep{potekhin.chabirer:eos_solid} based on recent Monte Carlo simulations \citep{Dewitt1999Screening-enhan,Caillol1999Thermodynamic-l}. As a result, our liquid-phase free-energy is slightly larger (by 0.005\% at $\Gamma = 175$) than the expression of \citet{potekhin.chabirer:eos_solid}, and our freezing point is slightly higher than the value of $\Gamma = 175\pm0.4$ determined by \citet{potekhin.chabirer:eos_solid}. Although we use the value of $\Gamma$ at the freezing point determined by the ion free energy for a one-component plasma (OCP), it is important to note that the freezing point may differ significantly from this value because of polarization of the electron background \citep{potekhin.chabirer:eos_solid} and because the crust consists of a mixture of isotopes. Indeed, recent molecular dynamics simulations \citep{Horowitz2007Phase-Separatio} find that the freezing temperature for a mixture appropriate for X-ray burst ashes is smaller than the OCP, with $\Gamma\approx 250$ at freezing.

We include the Debye reduction in the ion specific heat using an interpolation formula, and compute the Debye temperature according to \citet{chabrier93:_quant_coulum_applic}.  In the inner crust, we compute the neutron specific heat as that of a degenerate gas, with suppression due to the pairing interaction \citep{levenfish94}. For the $^{1}S_{0}$ pairing of free neutrons in the inner crust, we approximate the critical temperature as a Gaussian in the neutron Fermi wavevector, which approximates the calculation of \citet{Ainsworth1989superfluid-tc}. The crust temperatures achieved in the models presented here lie below the critical temperature throughout most of the crust, and the neutrons do not contribute significantly to the total specific heat.

The composition of accreting neutron star crusts has been calculated previously by \cite{Haensel2008Models-of-crust} and \cite{Gupta2006Heating-in-the-}. They find that the composition of the crust changes abruptly with depth, at locations corresponding to thresholds for electron capture or pycnonuclear reactions. Rather than model the neutron star crust as a series of distinct layers, we instead fit the composition so that $Z$ and $A$ are smoothly varying functions of $p$.  This approximation has a negligible effect on our results, but makes the integration simpler as it avoids jumps in the thermal properties. In the outer crust, we set the mass number $A$ and compute $Z$ in order to maintain $\beta$-equilibrium.  Using a simple liquid-drop model for the nuclear binding energy,
\begin{equation}\label{e.liquid-drop}
B(A,Z) = a_{V}A - a_{S}A^{2/3} - a_{A}\frac{(N-Z)^{2}}{A} - a_{C}\frac{Z^{2}}{A^{1/3}},
\end{equation}
and minimizing the nucleon-specific Gibbs free energy $G=Y_e \mu_{e}-B(A,Z)/A$  of a cell containing a nucleus and $Z$ electrons with respect to $Z$ for a fixed $A$,
we obtain the electron fraction $Y_{e}$ in the outer crust as a function of electron chemical potential $\mu_{e}$, 
\begin{equation}\label{e.Ye}
Y_{e} \approx \left(\frac{1}{2}- \frac{\mu_{e}}{8 a_{A}}\right)\left(1+\frac{a_{C}A^{2/3}}{4a_{A}}\right)^{-1},
\end{equation}
where we take\footnote{We obtain values for $a_A$ and $a_C$ by fitting experimental binding energies using the calculator at \url{http://128.95.95.61/\textasciitilde intuser/ld.html}.} $a_{A} = 23.43\nsp\MeV$ and $a_{C} = 0.715\nsp\MeV$. This formula accurately reproduces $Y_{e}$ in the crust models computed by \citet{Gupta2006Heating-in-the-}.

\begin{figure}[htbp]
\centering{\includegraphics[width=0.4\textheight]{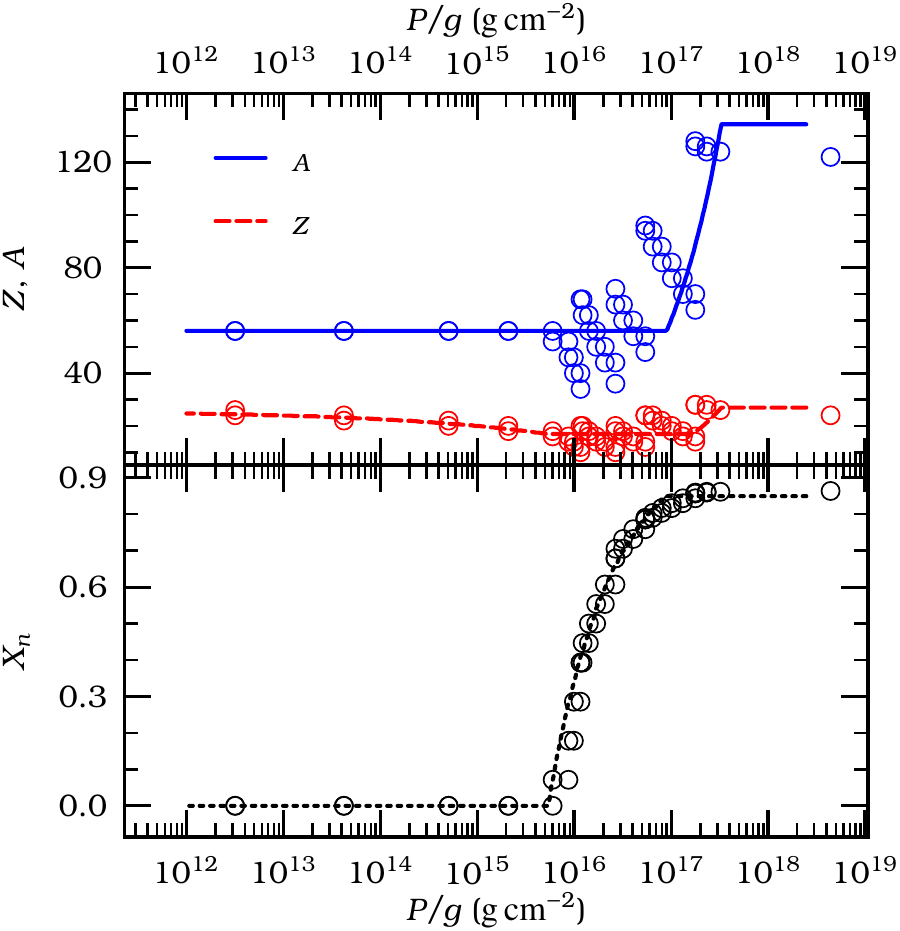}
\caption{\label{f.composition}Model of composition in the crust: nuclear charge (\emph{top panel, dashed line}) and mass (\emph{top panel, solid line}), and mass fraction of free neutrons (\emph{bottom panel}), as functions of depth coordinate $P/g$. The composition from \citet{Haensel2008Models-of-crust} is shown (\emph{open circles}) for comparison.}}
\end{figure}

In the inner crust, the relation between chemical potential, \Ye, and the abundance of free neutrons \YN\ becomes more complicated. We find that the tables in \citet{Haensel2008Models-of-crust} are adequately fit by defining $A_{\mathrm{tot}}$ as the total number of nucleons (including free neutrons) per nucleus and setting $Z/A_{\mathrm{tot}}\propto p^{-2/3}$ and $A_{\mathrm{tot}}$ increasing as $A_\mathrm{tot}\propto p^{2/3}$ up to a maximum value set by the total number of pycnonuclear reactions that occur in the crust. We set a floor to $Z/A_{\mathrm{tot}}$ of 0.03, and take $A = \max(A_{\mathrm{outer}},0.15 A_{\mathrm{tot}})$, where $A_{\mathrm{outer}}$ is the mean mass number in the outer crust. Figure \ref{f.composition} shows an example of a crust composition with $A_{\mathrm{outer}}=56$ and $A_{\mathrm{tot,max}} = 896$ in the inner crust, where we choose $A_{\mathrm{tot,max}}$ to agree with the calculation of  \citet[][Fig.~\ref{f.composition}, \emph{open circles}]{Haensel2008Models-of-crust}.

\subsection{Nuclear heating and neutrino cooling}\label{s.heating-cooling}

\begin{figure}[htbp]
\centering{\includegraphics[height=0.3\textheight]{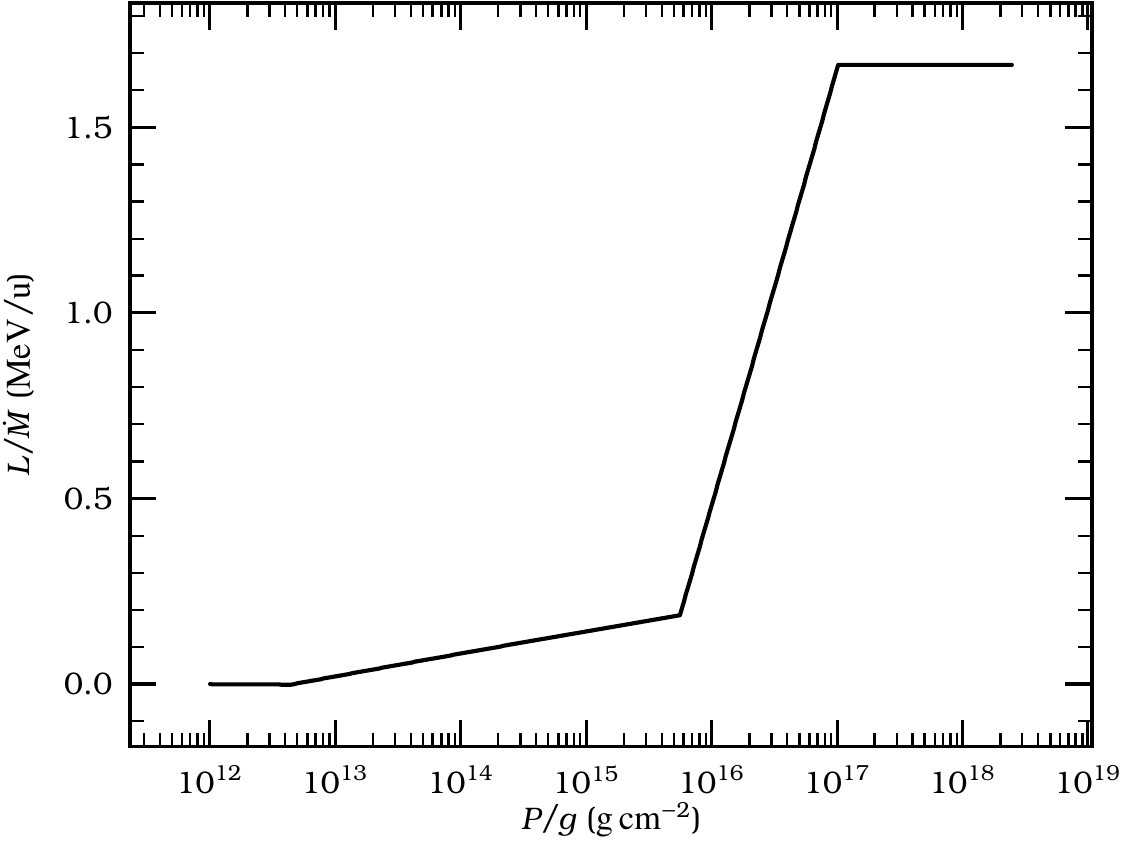}
\caption{Integrated nuclear heating, divided by the proper mass accretion rate, in the crust as a function of column.  \label{f.model-heating}}}
\end{figure}

Following \citet{brown:nuclear}, we define a smooth heating distribution in the crust, rather than resolving the heating from individual reaction layers. We choose our heating function to be such that $\dif L_{\mathrm{nuc}}/\dif\ln y = \textrm{const}$, and we do this separately in both the inner crust, and in the outer crust where the pressure is $P > 10^{27}\nsp\ergs\usp\cm^{-3}$. The integrated nuclear luminosity is plotted in Fig.~\ref{f.model-heating}.
We normalized the heat distribution so that the total heat deposited, per accreted nucleon, into the inner crust is $1.5\nsp\MeV$ \citep[cf.][]{haensel90a,haensel.zdunik:nuclear,Haensel2008Models-of-crust} and the total heat deposited, per accreted nucleon, into the  outer crust is $0.2\nsp\MeV$ \citep[cf.][]{Gupta2006Heating-in-the-}.

For the neutrino cooling, our model includes  \citep[for a review of neutrino emission mechanisms, see][]{yakovlev.kaminker.ea:neutrino} neutrino cooling from electron-nucleus bremsstrahlung.  The neutrino emissivity 
from neutrons paired in the $^{1}S_{0}$ state in the inner crust is suppressed by a factor $(v_{\mathrm{F}}/c)^{2}$ \citep{Leinson2006Vector-Current-,SedrakianVertex-renormal}. Recent calculations \citep{Steiner2009Superfluid-Resp} show that this suppression follows from conservation of baryon vector current. The pair, photo, and plasmon emissivities \citep{schinder87} do not contribute substantially at the temperatures of interest.

\subsection{Thermal conductivities}\label{s.thermal-conductivity}

Our implementation of the thermal conductivities mediated by electron-ion scattering follows that of \citet{potekhin99:_trans} and \citet{gnedin.ea:thermal_relaxation}. We compute the electron thermal transport in the relaxation-time approximation,
\begin{equation}\label{e.Wiedemann}
K = \frac{\pi^{2}}{3}\frac{n_{e}\kB^{2}T}{m_{e}^{\star}\nu},
\end{equation}
where $m_{e}^{\star} = (p_{F}^{2}/c^{2} + m_{e}^{2})^{1/2}$, with $p_{F}$ being the Fermi momentum, and $\nu$ is the scattering frequency. In the ocean, $\nu$ is set by electron-ion scattering. As the ions crystallize, electron-phonon scattering mediates the thermal transport. Where the temperature is above the Debye temperature, the scattering frequency is approximately
\begin{equation}\label{e.phonon-freq}
\nu_{ep} \approx 13\alpha \frac{\kB T}{\hbar},
\end{equation}
where $\alpha = e^{2}/(\hbar c)$ is the fine-structure constant.  In the inner crust, the electron-ion scattering frequency is strongly reduced for $T<T_{p}$, the plasma temperature, and impurity scattering becomes dominant with scattering frequency
\begin{equation}\label{e.impurity-freq}
\nu_{eQ} = \frac{4\pi \Qimp e^{4} n_{\mathrm{ion}}}{p_{F}^{2}v_{F}}\Lambda_{\mathrm{imp}},
\end{equation}
where $p_{F}$ and $v_{F}$ are the momentum and velocity of electrons at the Fermi surface and the impurity parameter $\Qimp \equiv n_{\mathrm{ion}}^{-1}\sum_{i}n_{i}(Z_{i}-\langle Z\rangle)^{2}$ measures the distribution of nuclide charge numbers. Although we do include electron-electron scattering \citep{urpin80:_therm,potekhin97} in our calculation, it does not affect the overall thermal conductivity.

For the Coulomb logarithm term $\Lambda_{\mathrm{imp}}$ we use the formula of \citet{potekhin99:_trans} with the modification that the structure factor is set to unity, reflecting the lack of long-range correlations in the impurities. With this modification $\Lambda_{\mathrm{imp}}$ becomes (Potekhin, private communication)
\begin{equation}\label{e.coulombL-eQ}
\Lambda_{\mathrm{imp}} = \frac{1}{2}\left\{\left[1+2\beta^{2}\frac{q_{s}^{2}}{2k_{\mathrm{F}}^{2}}\right] 
	\ln\left(1+\frac{4k_{\mathrm{F}}^{2}}{q_{s}^{2}}\right) - \beta^{2} - \frac{1+\beta^{2}(q_{s}/2k_{\mathrm{F}})^{2}}{1+(q_{s}/2k_{\mathrm{F}})^{2}}\right\},
\end{equation}
where $k_{\mathrm{F}} = p_{\mathrm{F}}/\hbar$ is the Fermi wavevector, $\beta = v_{\mathrm{F}}/c$ is the Fermi velocity, and $q_{s}\approx k_{\mathrm{TF}}$ is the Thomas-Fermi wavevector. As noted by \citet{brown.bildsten.ea:variability}, this gives a result for the Coulomb logarithm that is similar to that proposed by \citet{itoh93}. The two formulae agree if one makes the substitution $k_{\mathrm{F}}a\to k_{\mathrm{F}}/(k_{\mathrm{TF}}/2)$, where $a$ is the mean inter-nuclei spacing, in the fit by \citet{itoh93}. The Thomas-Fermi screening length exceeds the inter-nuclei spacing and produces a larger $\Lambda_{\mathrm{imp}}$ (by about a factor of 2) and hence a lower thermal conductivity than does the fit by \citet{itoh93}. Molecular dynamics simulations \citep{Horowitz2008Thermal-conduct} using a mixture of rp-process ashes \citep{Gupta2006Heating-in-the-}, find that the impurities are not distributed uniformly. As a result, \citet{Horowitz2008Thermal-conduct} compute a thermal conductivity that is $\approx 30\%$ lower than that computed assuming a one-component plasma and impurities computed with the static structure factor of \citet{itoh93}. 

Finally, \cite{Aguilera2008Superfluid-Heat} recently calculated the thermal conductivity in the inner crust due to superfluid heat conduction. Their Figure 2 shows that, for $T = 10^{8}\nsp\K$, superfluid heat conduction is comparable to heat conduction by electrons for $\rho_{\mathrm{nd}} < \rho \lesssim 2\rho_{\mathrm{nd}}$, where $\rho_{\mathrm{nd}}$ is the neutron drip density, and less important at lower temperatures. Therefore, we do not expect that our results would change significantly if superfluid heat conduction was included, unlike the case of magnetars, for which the strong magnetic field suppresses electron transport perpendicular to the magnetic field. 

\end{document}